\documentclass[aps,superscriptaddress,amsmath,amssymb,preprint]{revtex4}
\usepackage{amsmath,amsfonts,amssymb}
\usepackage{graphicx,subfigure}
\usepackage{epsfig}

\begin{document}

\title{Self-trapping and acceleration of ions 
in laser-driven relativistically transparent plasma
}

\date{8 Oct 2018}

\author{B. Liu}
\affiliation{Institute for Computational and Plasma Physics, Ludwig-Maximilian-Universitaet, Muenchen, 80333 Muenchen, Germany}

\author{J. Meyer-ter-Vehn}
\affiliation{Max-Planck-Institut f\"{u}r Quantenoptik, D-85748 Garching, Germany}

\author{H. Ruhl}
\affiliation{Institute for Computational and Plasma Physics, Ludwig-Maximilian-Universitaet, Muenchen, 80333 Muenchen, Germany}

\begin{abstract}
Self-trapping and acceleration of ions 
in laser-driven relativistically transparent plasma
are investigated with the help of particle-in-cell simulations. 
A theoretical model based on ion wave breaking is established 
in describing ion evolution and ion trapping. 
The threshold for ion trapping is identified. 
Near the threshold 
ion trapping is self-regulating and stops 
when the number of trapped ions is large enough.
The model is applied to ion trapping 
in three-dimensional geometry. 
Longitudinal distributions of ions 
and the electric field near the wave breaking point 
are derived analytically in terms of power-law scalings. 
The areal density of trapped charge is obtained as a function of 
the strength of ion wave breaking, which scales with 
target density for fixed laser intensity.  
The results of the model are confirmed by the simulations.
\end{abstract}

\maketitle

\section{Introduction}

Laser-driven plasma-based ion acceleration 
has attracted substantial interest
due to its potential applications in many areas \cite{rev1, app1, app2, app3},  
including medical treatment of cancer, matter detection, 
fast ignition in inertial confinement fusion, 
nuclear physics, and high energy physics.

Recent developments of ultra-intense  
laser technology have opened new options for  
generating high energy ion beams \cite{hb,rpa,piston,robinson,weng,rita,sasl,sasl3,iwba}.
An ultra-intense laser pulse is capable of producing a  
localized charge-separation field in the region behind of laser front. 
Ions trapped in the field are accelerated to velocities far beyond 
the velocity of the laser front.  
In order to achieve high energy ion acceleration, it is necessary to 
make the accelerating field propagate as fast as possible. 
However, when it is too fast background ions cannot be self-trapped. 
There must exist a threshold condition where ion trapping initiates.   
For a given laser pulse
the most energetic ions are produced under the threshold condition.

Here we focus on ion trapping and acceleration 
near the threshold condition.  
This has not been studied in sufficient detail so far. 
In previous work \cite{iwba} 
we have pointed out  
that ion trapping near the threshold condition
is caused by ion wave breaking. 
In the present paper, 
with the help of particle-in-cell (PIC) simulations, 
we develop a theoretical model to investigate the ion trapping
in more detail. 
We demonstrate that the ion trapping initiates with 
the transition from fluid-like to kinetic behaviour, 
accompanied by the transition from oscillatory to self-trapping dynamics, 
as well as the transition from non-crossing to crossing trajectories. 
It happens in near-critical \cite{ncdt1,ncdt2,ncdt3,ncdt4} 
relativistically transparent \cite{trans} plasma. 
The interesting point is that this trapping process 
is self-regulating and stops when the number of trapped ions is large enough. 
This is significantly different from 
ion acceleration regimes in opaque plasma, 
such as hole boring \cite{hb}, laser piston \cite{piston}, 
and radiation pressure acceleration for thin foils \cite{rpa},
where all ions in the laser focuses are accelerated.  
This results in high-quality ion beams  
with low energy spread and beam emittance. 
The number of trapped ions 
can be controlled by external parameters
like laser intensity and target density.
It allows to design robust and controllable 
laser-plasma ion accelerators.

A few efforts have been made previously on 
ion acceleration in laser-driven plasma waves or wave-like structures 
in near-critical relativistically transparent plasma. 
In Ref. \cite{ionwake}, O. Shorokhov and A. Pukhov have proposed 
the so-called ion wakefield acceleration mechanism 
in which ions are trapped in an electron plasma wave and accelerated. 
Ion acceleration in bubble regime in 3D geometry has been investigated in Ref. \cite{ionbub}. 
Ion acceleration in laser-driven comoving electrostatic field 
in inhomogeneous plasma has been studied  
in the scheme of relativistically induced transparency acceleration \cite{rita}.  
In all these works, the fields that trap and accelerate ions 
are produced by the oscillations of electrons.
In order to maintain the fields, stable and positively charged backgrounds are required. 
Two-component plasma targets are used in these works 
where a large proportion of heavier (larger mass-to-charge ratio) ions 
compose the background and a small proportion of 
lighter ions are trapped and accelerated.  
This makes a high demand on target preparing and 
limits their applications. 
In our work, background ions are significantly disturbed and 
ion waves are excited. 
The ion wave survives even though an electron wave is completely destroyed. 
Therefore, ion acceleration via ion wave breaking works well 
even when the plasma is composed of only one kind of ions.

The present paper is organized as follows. 
In Sec. II, we investigate ion trapping via ion wave breaking
based on a 1D model. 
In Sec. III, we apply the 1D model to 
estimate ion trapping 
in practical 3D geometry. 
In Sec. IV, we compare the model predictions with 
3D PIC simulations.  
Conclusions are given in Sec. V.

\section{1D model of ion wave breaking 
}

\subsection{Electron wave and ion wave}

We follow the pioneering work of A. I. Akhiezer and R. V. Polovin 
\cite{Akhiezer}. 
We consider a one-dimensional plane  plasma wave 
(longitudinal oscillation)
propagating in an uniform 
unmagnetized 
plasma along $z$-direction
with 
phase velocity $v_{\rm{ph}}$. 
The difference here is that we take ion motion into account. 
We assume the plasma is composed of electrons 
and one kind of ions, having number densities
$n_e$ and $n_i$, respectively. 
The plasma is described 
by the equations of motion and continuity 
for electrons, 
\begin{eqnarray}
\left(\frac{\partial }{\partial t}+ v_e\frac{\partial }{\partial z}\right) (m_e \gamma_e  v_e )&=&  -e E_z, 
\label{ele_momentum}  \\
\frac{\partial n_e }{ \partial t} +
\frac{ \partial}{\partial z } (n_e v_e) &=& 0 ,
\label{ele_continuity}
\end{eqnarray} 
and ions, 
\begin{eqnarray}
\left(\frac{\partial }{\partial t}+ v_i\frac{\partial }{\partial z}\right) (m_i \gamma_i  v_i )&=&  q_i E_z, 
\label{ion_momentum}    \\
\frac{\partial n_i }{ \partial t} +
\frac{ \partial}{\partial z } (n_i v_i) &=& 0 ,
\label{ion_continuity}
\end{eqnarray} 
respectively, 
and Poisson's equation \cite{footnote0}, 
\begin{eqnarray}
\frac{\partial}{\partial z }E_z &=& \frac{1}{\epsilon_0} (q_{i} n_{i}-e n_e),
\label{poisson}
\end{eqnarray} 
where $\gamma_{e,i}=1/\sqrt{1-\beta_{e,i}^2}$, $\beta_{e,i}=v_{e,i}/c$,
$v_e$ ($v_i$) denotes the electron (ion) velocity,
$c$ the light speed in vacuum, 
$e$ ($q_i$) the electron (ion) charge, 
$m_e$ ($m_i$) the electron (ion) mass,  
$E_z$  the longitudinal electric field, 
and $\epsilon_0$ the vacuum permittivity.

If the velocity $v_{\rm{ph}}$ is constant, 
all the variables can be written as 
functions of 
$\zeta=\omega_i (z/v_{\rm{ph}}-t)$ 
(see Ref. \cite{Akhiezer}), 
instead of $z$ and $t$ separately, 
where $\omega_i=\sqrt{q_i e n_{0}/m_i \epsilon_0}$ 
is the ion oscillation frequency, 
$e$ denotes the electron charge, 
and $n_{0}$  the initial plasma density. 
We look for stationary wave solutions 
$n_{e,i}(\xi)$, $\beta_{e,i}(\xi)$, and $E_z(\xi)$. 
Introducing dimensionless quantities
$\tilde{n}_e= n_e/ n_{0},$
$\tilde{n}_i=q_i n_i/ e n_{0},$
and 
$\tilde{E}=E_z/E_0$,
where  
$E_0=m_i  c \omega_i/q_i$,  
Eqs. (\ref{ele_momentum}-\ref{poisson}) transform into
\begin{eqnarray}
(1-\beta_e/\beta_{\rm{ph}})\frac{d \gamma_e \beta_e}{d\zeta}&=& \mu \tilde{E},
\label{eqbetae}  \\
\frac{d }{d \zeta} \left( \tilde{n}_e  - 
\tilde{n}_e  \beta_e/\beta_{\rm{ph}} \right) &=& 0, 
\label{eqnne}    \\
(1-\beta_i/\beta_{\rm{ph}})\frac{d \gamma_i \beta_i}{d\zeta}&=&  -\tilde{E},
\label{eqbetai0}  \\
\frac{d }{d \zeta} \left(  \tilde{n}_i  - 
\tilde{n}_i  \beta_i/\beta_{\rm{ph}} \right) &=& 0,
\label{eqnni}     \\ 
\frac{d  \tilde{E}}{d \zeta} &=&  \beta_{\rm{ph}}   (\tilde{n}_i - \tilde{n}_e ),   
\label{eqez0}      
\end{eqnarray}
where $\beta_{\rm{ph}} = v_{\rm{ph}}/c$,  and 
$ \mu = (m_i/m_e)/(q_i/e)$. 
The densities, 
\begin{eqnarray}
\tilde{n}_e &=& \frac{1}{1-\beta_e/\beta_{\rm{ph}}}, 
\label{eqne} \\
\tilde{n}_i &=& \frac{1}{1-\beta_i/\beta_{\rm{ph}}}, 
\label{eqni0} 
\end{eqnarray}
are obtained directly from Eqs. (\ref{eqnne}, \ref{eqnni}) 
since $\tilde{n}_e \equiv \tilde{n}_i \equiv 1$ 
for unperturbed plasma 
($\beta_e \equiv \beta_i \equiv 0$).

\begin{figure}[htb]
\centering
\includegraphics[width=0.85\textwidth]{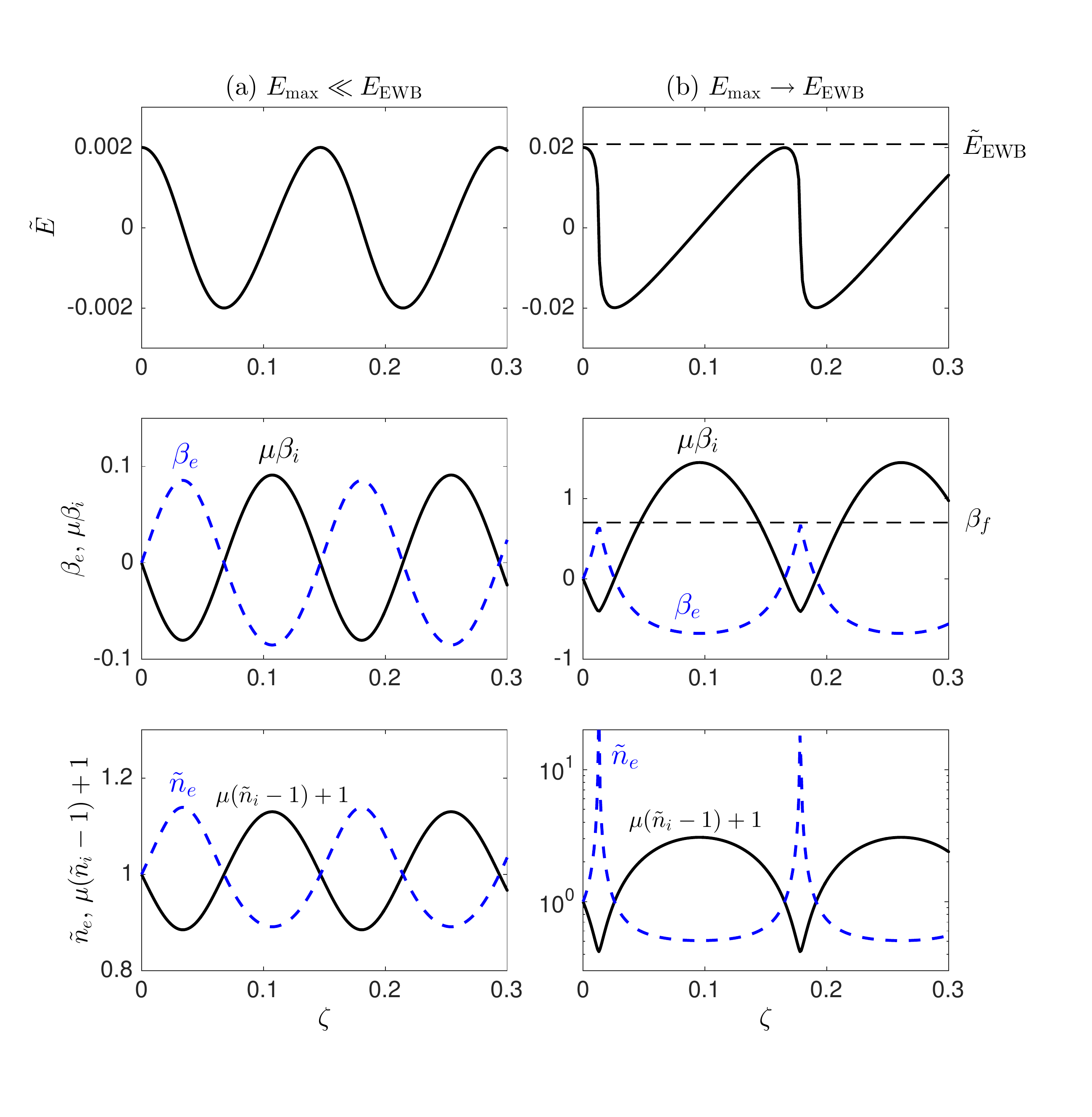}
\caption{
Results of (top panels) 
longitudinal electric field $\tilde{E}$, 
(middle panels) electron (ion) velocity $\beta_e$ ($\beta_i$)  
and (lower panels) density $\tilde{n}_e$ ($\tilde{n}_i$),
by numerically solving Eqs. (\ref{eqbetae}, \ref{eqbetai0}, 
\ref{eqez0}, \ref{eqne}, \ref{eqni0}) 
with $\beta_{\rm{ph}} = 0.7$, and 
(a) $E_{\rm{max}} = 0.002$,  
(b) $E_{\rm{max}}=0.020$,
where $\mu = 1836$ for hydrogen plasma. 
The electron wave breaking field $E_{\rm{EWB}} = 0.021$ is 
calculated according to Eq. (\ref{eewb}). 
The electron wave is accompanied by an ion wave.
\label{ana1d1}
}
\end{figure}

By numerically solving 
Eqs. (\ref{eqbetae}, \ref{eqbetai0}, \ref{eqez0},
\ref{eqne}, \ref{eqni0}) 
with $\beta_{\rm{ph}}=0.7$ and boundary conditions 
$\beta_e(0)=0$, $\beta_i(0)=0$, 
and $\tilde{E}_z(0)=0.002$,  
the distributions of electrons, ions, 
and the longitudinal electric field 
are shown in Fig. \ref{ana1d1} (a).   
It is shown that the electron wave is accompanied by an ion wave. 
The electron wave, the ion wave, and the longitudinal electric field 
are all oscillating sinusoidally.  
The ion wave peaks at the electron wave valley 
where $E_z$ changes sign from negative to positive.
As $\mu \gg 1$, the amplitude of the ion wave is far smaller 
than that of the electron wave.

As is well-known \cite{Akhiezer,dawson,ewbrev}, 
the amplitude of plasma wave can be characterized by the maximum longitudinal
electric field $E_{\rm{max}}$, 
and there exists a threshold field for 
electron wave breaking (here we denote it by $E_{\rm{EWB}}$), 
beyond which the fluid-like behaviour of electrons breaks down
and there is no solution for Eqs. (\ref{eqbetae}, \ref{eqbetai0}, \ref{eqez0},
\ref{eqne}, \ref{eqni0}). 
When $\mu \gg 1$, 
the threshold field in Ref. \cite{Akhiezer} 
is reproduced \cite{footnote1} as
\begin{eqnarray}
\tilde{E}_{\rm{EWB}} = \sqrt{2(\gamma_{\rm{ph}}-1)/\mu}, 
\label{eewb}
\end{eqnarray}
where $\gamma_{\rm{ph}}=1/\sqrt{1-\beta_{\rm{ph}}^2}$. 
Figure \ref{ana1d1} (b) shows the results 
when $E_{\rm{max}}$ approaches $E_{\rm{EWB}}$. 
As is also well-known,  
the electron wave becomes strongly nonlinear, and the electric field 
becomes a sawtooth-like wave. 
The interesting thing is that 
the ion wave is also strongly nonlinear, instead of sinusoidal-like, 
although its amplitude is still very small.

\subsection{Ion wave breaking}
\label{seciwb}

\subsubsection{Equations of ions}

A plasma wave can be produced by propagating a laser pulse in plasma. 
In 1D geometry, the electric field and plasma along the laser propagation 
direction is well described by 
Eqs. (\ref{eqbetae},\ref{eqbetai0},\ref{eqez0},\ref{eqne},\ref{eqni0})
when the maximum electric field satisfies $E_{\rm{max}} < E_{\rm{EWB}}$. 
When the laser drive is strong enough
so that $E_{\rm{max}} > E_{\rm{EWB}}$, 
self-trapping of electrons happens, and 
the fluid model for electrons breaks down. 
However, ions still behave like a fluid and 
Eqs. (\ref{ion_momentum},\ref{ion_continuity},\ref{poisson}) 
are still valid, 
as long as there is no self-trapping of ions.

We denote the propagating velocity of the laser front by $v_f$. 
If $v_f$ is constant, and the plasma is stationary in the frame comoving 
with $v_f$, then, similar to that in Sec. II A, 
all the variables can be written as 
functions of $\xi=\omega_i (z/v_f-t)$, 
and Eqs. (\ref{ion_momentum},\ref{ion_continuity},\ref{poisson}) 
become, 
\begin{eqnarray}
(1-\beta_i/\beta_{f})\frac{d \gamma_i \beta_i}{d\xi}&=&  -\tilde{E},
\label{eqbetai}  \\
\tilde{n}_i &=& \frac{1}{1-\beta_i/\beta_{f}}, 
\label{eqni}     \\
\frac{d  \tilde{E}}{d \xi}&=&  \beta_{f}   (\tilde{n}_i - \tilde{n}_e ),
\label{eqez}
\end{eqnarray}
where $\beta_f=v_f/c$.

\subsubsection{1D PIC simulation}

In order to investigate the general features of ion evolution 
under the condition of $E_{\rm{max}}>E_{\rm{EWB}}$, 
a 1D PIC simulation has been carried out 
by using the plasma simulation code (PSC) \cite{psc}.
The simulation makes use of a constant laser pulse with 
circular polarization
(similar features exist for linearly polarized cases)
and laser intensity  
$I_L=5\times 10^{21}\rm{W/cm^2}$ 
(laser amplitude $a_0=44$ for wavelength $1\rm{\mu m}$, 
where $a_0=e E_L/m_e c \omega_L$, $E_L$ denotes the laser electric field, 
and $\omega_L$ the laser frequency)
of semi-infinite duration, and 
a sine-squared front edge rising over 5 laser periods. 
The initial plasma density rises linearly from $z=5\mu m$ to $z=6\mu m$ 
and then keeps constant as 
$n_0=0.2n_c$ until the end of the simulation box at $z=100\mu m$,
where $n_c = m_e \omega_L^2 \epsilon_0 /e^2$ is the critical plasma density. 
The plasma is composed of electrons and protons. 
A resolution of 500 cells per micron and 20 particles per species 
per cell is used. 
The initial temperature is 100 eV for electrons and 0 for ions.

\begin{figure}[htb]
\centering
\includegraphics[width=0.45\textwidth]{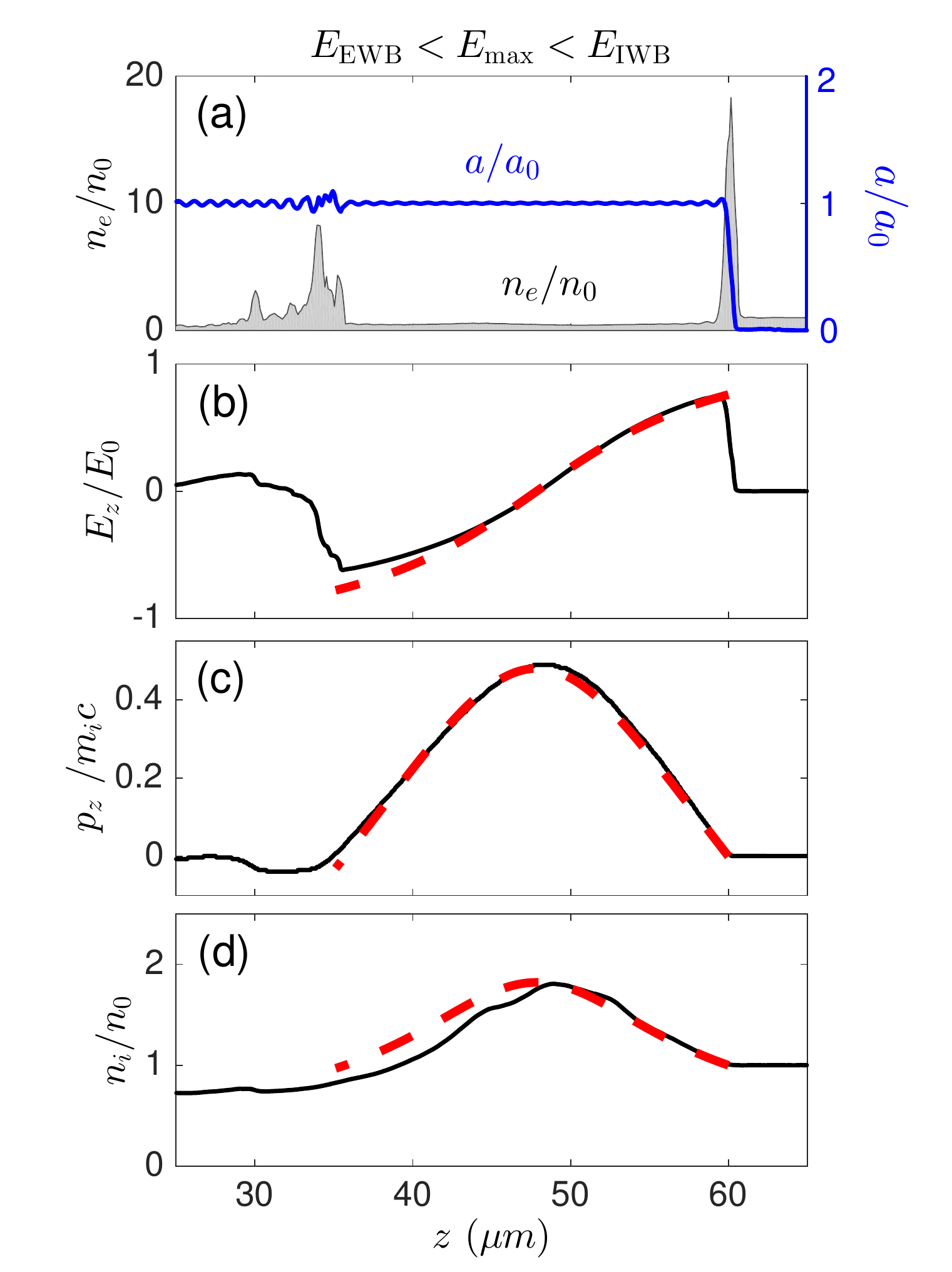}
\caption{
A snapshot of (a) laser amplitude ($a/a_0$, blue line), 
electron density ($\tilde{n}_e$, grey area), 
(b) electric field ($\tilde{E_z}$, black solid line), 
(c) ion momentum ($p_z/m_i c$, black solid line), 
and (d) ion density ($\tilde{n}_i$, black solid line), 
from a 1D PIC simulation 
with the laser amplitude $a_0=44$ and 
initial plasma density $n_0=0.2n_c$. 
An ion wave survives  
even though an electron wave does not exist.
The results of the model (red dashed lines) 
are obtained by numerically solving 
Eqs. (\ref{eqbetai}-\ref{eleden})  
with quantities  
$\beta_f \sim 0.96$, $v_b \sim v_i (z=60\mu m) \sim 0$, 
$E_{\rm{max}} \sim E_z(z=60\mu m) \sim 0.75 E_0$, 
and 
$\tilde{n}_{el} \sim 0.5$ in the range of $35\mu m<z<60\mu m$,
observed in the simulation.  
The results are %show an ion wave 
under the condition of $E_{\rm{EWB}}<E_{\rm{max}}<E_{\rm{IWB}}$, 
where $E_{\rm{EWB}}=0.05 E_0$ is calculated
according to Eq. (\ref{eewb}) by assuming $\beta_{\rm{ph}}=\beta_f$, 
while $E_{\rm{IWB}}= 2.42 E_0$ is calculated
according to Eq. (\ref{efthreshold}). 
\label{sim1d1}
}
\end{figure}

\begin{figure}[htb]
\centering
\includegraphics[width=0.45\textwidth]{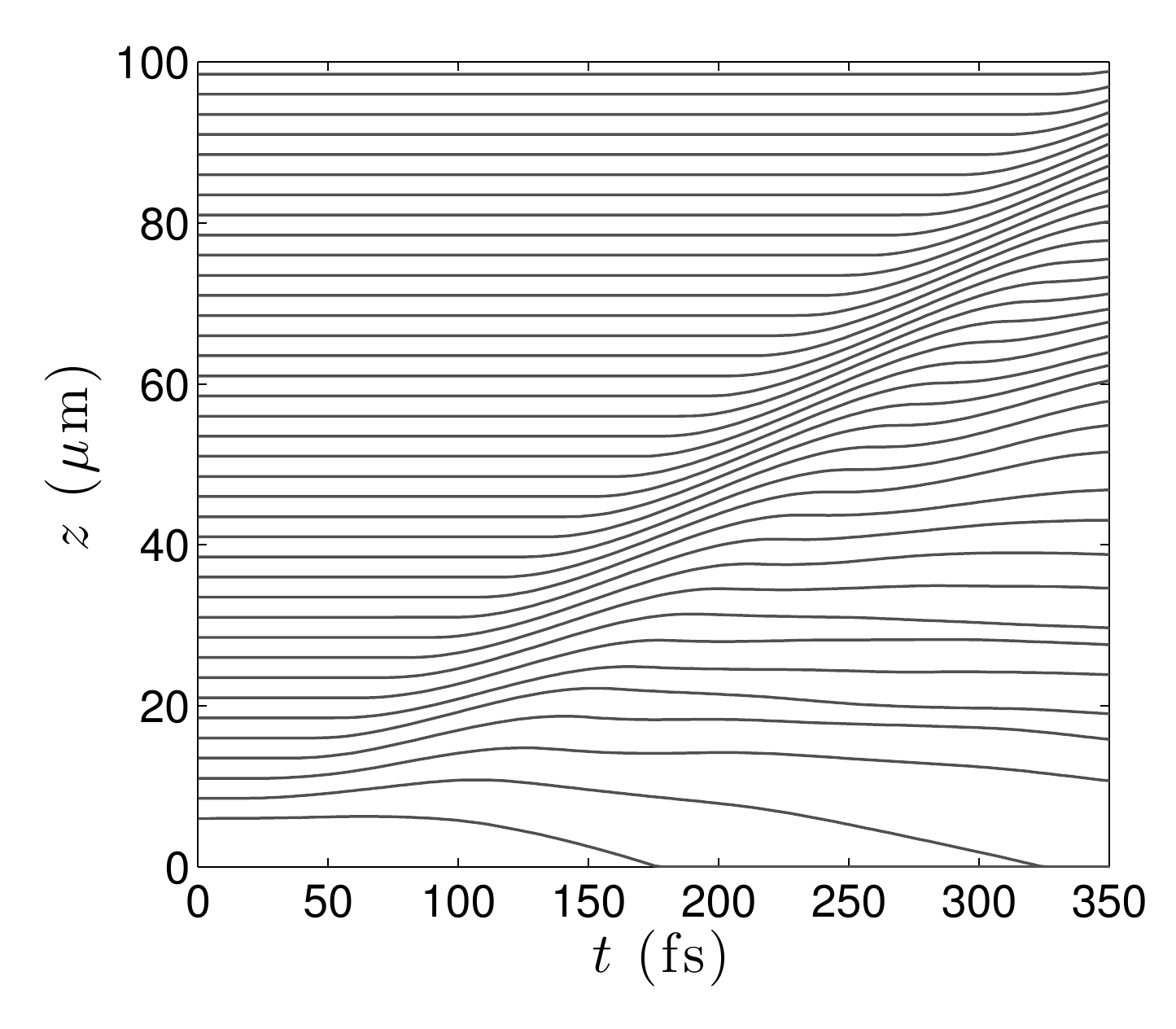}
\caption{Trajectories of ions with  different initial positions 
for the 1D simulation shown in Fig. \ref{sim1d1}.  
Ions show fluid-like behaviour. 
\label{traj1d1}
}
\end{figure}

The simulation results are shown in Fig. \ref{sim1d1}. 
The laser pulse pushes electrons and piles them up at the laser front, 
forming a leading electron layer,   
and leaving a region behind of the laser front 
with relatively low electron density 
that is almost constant in the range of $35\mu m < z < 60\mu m$. 
A charge-separation field $E_z$ is then produced. 
Ions first acquire velocity in laser direction 
in the region where $E_z>0$,
but then loose it again in the region where $E_z<0$ (see Fig. \ref{sim1d1}(b,c)). 
Profiles of ion velocity and density are similar to that 
Fig. \ref{ana1d1} (b), when the electron wave is strongly nonlinear, 
except that in Fig. \ref{sim1d1} 
the oscillation lasts only one period and the amplitudes are much larger. 
Furthermore, trajectories of ions at different initial positions 
are plotted in Fig. \ref{traj1d1}. 
It is clearly shown that  
ions still behave like a fluid and there is no trajectory crossing.
In this sense, we can say that the ion wave survives in 
the region just behind of the laser front.

\subsubsection{Electron density}

\begin{figure}[htb]
\centering
\includegraphics[width=0.4\textwidth]{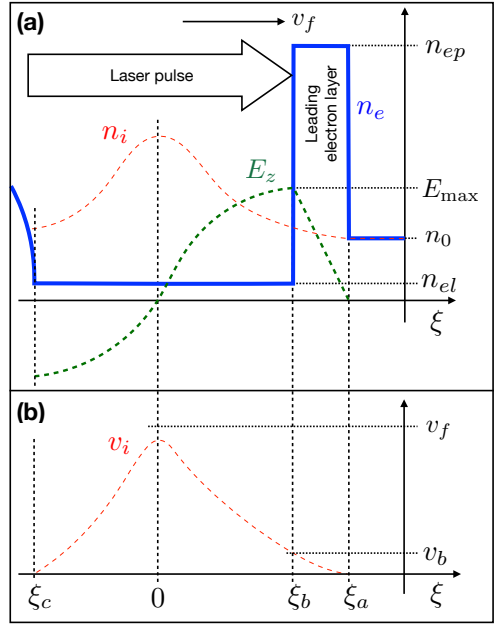}
\caption{
Schematic figure of ion dynamics under the condition of
$E_{\rm{EWB}}<E_{\rm{max}}<E_{\rm{IWB}}$, 
modelling the simulation results of Fig. \ref{sim1d1}. 
Plot (a): 
electron density (thick blue line), ion density (thin red line),
and electric field (green dashed line). 
Plot (b): ion velocity (thin red line). 
The laser front propagates forward with front velocity $v_f$.
The electron density distribution is modelled as 
$n_{ep}$ for $\xi_b<\xi<\xi_a$ and $n_{el}$ for 
$\xi_c<\xi<\xi_b$,  
where $\xi_c$ satisfies $v_i(\xi_c)=0$ and $E_z(\xi_c)<0$.
The electric field $E_z$ peaks at $\xi_b$, i.e., $E_{\rm{max}}=E_z(\xi_b)$). 
The ion velocity at $\xi_b$ is smaller than the laser front velocity 
($v_b < v_f$). 
We chose $\xi=0$ as the point where ion velocity and density peak
and $E_z$ changes sign. 
\label{mod1d}
}
\end{figure}

The ion wave is coupled with the electron density $\tilde{n}_e$ 
via Poisson's equation (Eq. (\ref{eqez})). 
Since the fluid behaviour of electrons breaks down, 
$\tilde{n}_e$ cannot be obtained from fluid equations now. 
Fortunately, ions are much heavier than electrons.  
Thus, ions evolve slowly compared to electrons 
and ion motion is insensitive to the fast variations 
of the electron density. 
Therefore, for simplicity, 
we approximate the electron density as a step function,
\begin{eqnarray}
{n}_e = 
\begin{cases}
  n_0, & \xi>\xi_a; \\
  {n}_{ep}, & \xi_b<\xi<\xi_a; \\
 {n}_{el}, & \xi_c<\xi<\xi_b,   \\
 \rm{not~of~concern}, & \xi<\xi_c, 
\end{cases}
\label{eleden}
\end{eqnarray}
where $\xi_c$ is chosen as the point where ions are decelerated to zero velocity
for the first time ($\beta_i(\xi_c)=0$ and $E_z(\xi_c)<0$),
$\xi_a$ and $\xi_b$ denote the boundaries of the leading electron layer. 
The modelled electron density is illustrated in Fig. \ref{mod1d}. 
Then the charge-separation field peaks at $\xi_b$, 
i.e., $E(\xi_b)=E_{\rm{max}}$. 
The ion velocity at $\xi_b$ is denoted by $v_b$. 
It is smaller than the laser front velocity ($v_b < v_f$). 
For simplicity, we chose $\xi=0$ as the point 
where the ion velocity peaks. 
Then, at the point $\xi=0$, the ion density also peaks 
(see Eq. (\ref{eqni}) and the electric field changes sign 
from negative to positive 
(according to Eq. (\ref{eqbetai}) and Eq. (\ref{eqez}), 
$\tilde{E}(0)=0$ and $d\tilde{E}/d\xi |_{\xi=0} > 0$).
For $\xi<\xi_c$, the electron density is too complex but  
not important for ion trapping and acceleration.
Thus, it is not of concern in our model.
Higher order modifications of the electron density 
may give more details. Nevertheless,  
this approximation works very well, especially when $n_{el} \ll n_i$.

%\subsubsection{Comparing model with simulation}

Results of the model of the electric field $E_z$, 
the ions momentum $p_z$, and the ion density $n_i$ 
are shown in Fig. \ref{sim1d1} (b,c,d), denoted by red dashed lines. 
It is seen that the curves from the model 
with parameters observed in the simulation 
fit the simulation results very well.

If there exists an electron wave, as is well-known, 
the phase velocity of the electron wave 
should equal the laser front velocity, i.e., 
$\beta_{\rm{ph}}=\beta_f \sim 0.96$.  
However, in this case, according to Eq. (\ref{eewb}), 
one has $\tilde{E}_{\rm{EWB}} = 0.05 < \tilde{E}_{\rm{max}}$. 
Thus, there is no electron wave anymore.  
This confirms that an ion wave can survive 
even though an electron wave does not exist.

\subsubsection{Ion wave breaking threshold}
\label{iwbth}

Similar to Eq. (\ref{eewb}),  
there exists a threshold field for ion wave breaking as well. 
Combining  Eqs. (\ref{eqbetai},\ref{eqni},\ref{eqez}), 
we obtain   
\begin{equation}
\frac{d}{d \xi}\left(\frac{ \tilde{E}^2}{2}
+  \beta_{ f  } \gamma_i \beta_i 
+ \tilde{n}_{e} \gamma_i (1 - \beta_f \beta_i )
\right) =0. 
\label{int1}
\end{equation}
Integrating Eq. (\ref{int1}) from $\xi=0$ to $\xi=\xi_b$, 
we find the maximum electric field,  
\begin{eqnarray}
\tilde{E}_{\rm{max}} = 
\sqrt{ 2 \beta_f(1-\tilde{n}_{el})(\gamma_{i}(0) \beta_{i}(0) - \gamma_b \beta_b) +2 \tilde{n}_{el} (\gamma_{i}(0)-\gamma_b)
}, 
\label{eqezez}
\end{eqnarray}
where $\beta_b=v_b/c$, and $\gamma_{b}=1/\sqrt{1-\beta_{b}^2}$. 
When the ion velocity peak $\beta_{i}(0)$
equals the velocity of the laser front $\beta_f$, 
the maximum electric field $E_{\rm{max}}$ marks the threshold of 
ion wave breaking, 
\begin{equation}
\tilde{E}_{\rm{IWB}}
=\beta_f \sqrt{2\gamma_f(1-\delta)}, 
\label{efthreshold}
\end{equation}
where  
$$
\delta = \frac{\tilde{n}_{el}}{\beta_f^2\gamma_f^2}
(\gamma_b \gamma_f (1-\beta_b\beta_f)-1)
+ \frac{\beta_b \gamma_b}{\beta_f \gamma_f},  
$$
and $\gamma_f = 1/\sqrt{1-\beta_f^2}$.

When $\tilde{E}_{\rm{max}} > \tilde{E}_{\rm{IWB}}$, incoming ions   
are accelerated to velocities equal to the laser 
front velocity at points $\xi>0$, 
where the electric field is positive ($E_z >0$). 
Then these ions will be further accelerated. 
In the frame comoving with the laser front, 
they are reflected. 
This is a process of self-injection of ions into the 
accelerating field. Then the ions are trapped and further 
accelerated. 
In this condition, the fluid-like behaviour of ions breaks down, 
and there is no solution of Eqs. (\ref{eqbetai},\ref{eqni},\ref{eqez}).

For the simulation shown in Fig. \ref{sim1d1}, with  
$\beta_f \sim 0.96$, $\tilde{n}_{el} \sim 0.5$, and $\beta_b \sim 0$,  
one has the threshold field $E_{\rm{IWB}} = 2.4 E_0$. 
While the observed maximum charge-separation field in the simulation is
$E_{\rm{max}} \sim 0.75 E_0$. 
It is still smaller than the threshold value 
($E_{\rm{max}} < E_{\rm{IWB}}$). 
This explains the fluid-like behaviour of ions 
in the simulation, 
as is clearly seen in Figs. (\ref{sim1d1},\ref{traj1d1}).

\begin{figure}[htb]
\centering
\includegraphics[width=0.85\textwidth]{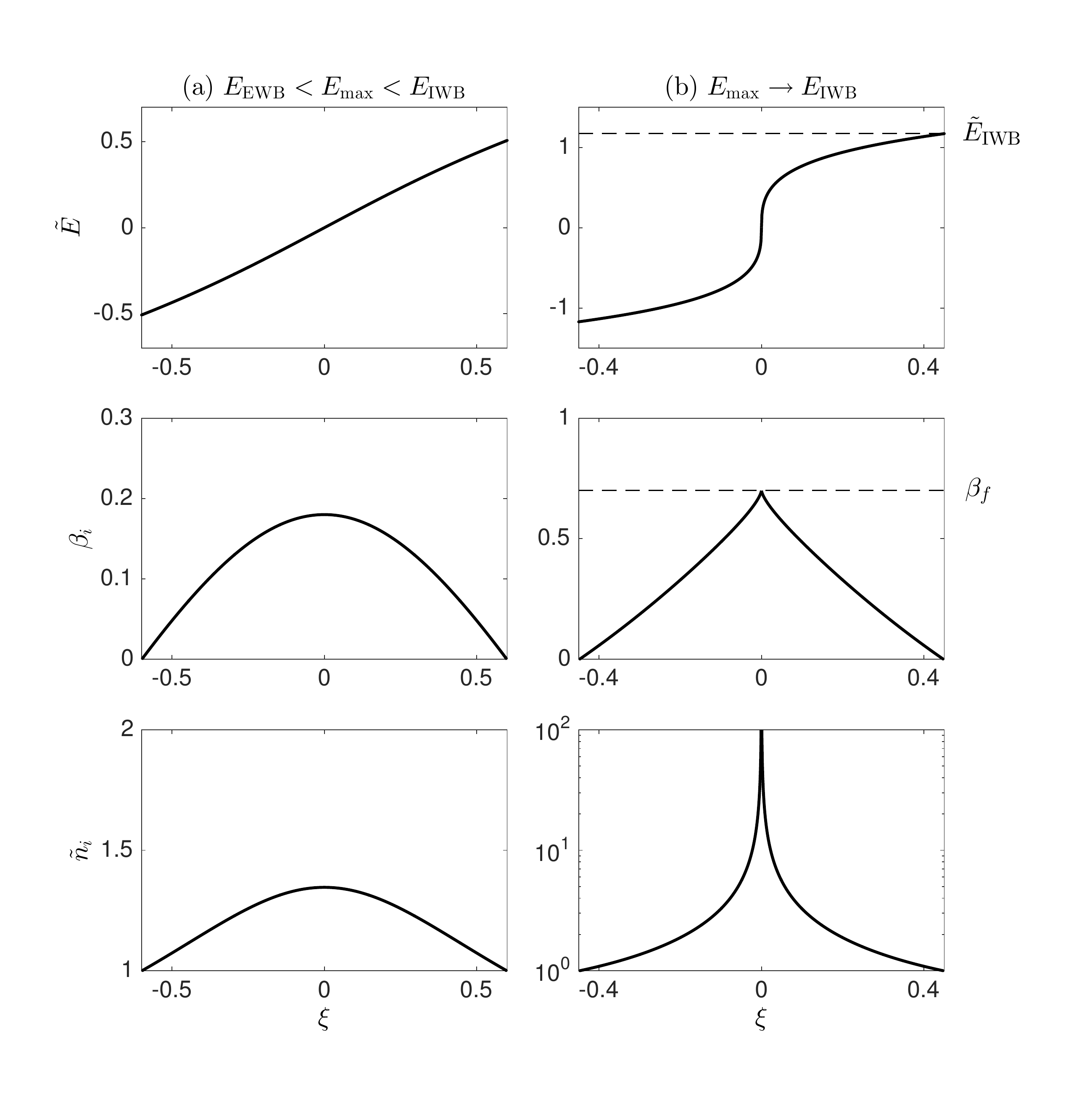}
\caption{Column (a):
Results of the model of (top panels) electric field $\tilde{E}$, 
(middle panels) ion velocity $\beta_i$, and (lower panels) ion density $\tilde{n}_i$, 
by numerically solving Eqs. (\ref{eqbetai}-\ref{eleden})  
with $\tilde{E}(0)=0$, $\beta_i(0)=0.18$, 
and the parameters $\beta_f = 0.7$, 
$n_{el}=0$, $v_b=0$. 
Column (b): The same as those in (a) except $\beta_i(0)\to \beta_f$.  
The results are under the condition of 
$\tilde{E}_{\rm{EWB}}<\tilde{E}_{\rm{max}}\leq \tilde{E}_{\rm{IWB}}$, 
where $\tilde{E}_{\rm{EWB}} = 0.02$ 
is calculated according to Eq. (\ref{eewb}) by assuming $\beta_{\rm{ph}}=\beta_f$, 
while $\tilde{E}_{\rm{IWB}} = 1.2$ is calculated according to
Eq. (\ref{efthreshold}).       
\label{ana1d2}
}
\end{figure}

Numerical solutions of  
Eqs. (\ref{eqbetai},\ref{eqni},\ref{eqez},\ref{eleden}) 
with different $E_{\rm{max}}$ 
are shown in Fig. \ref{ana1d2}. 
A weak ion wave is seen in Fig. \ref{ana1d2} (a), 
in which $E_{\rm{max}}$ is far smaller than $E_{\rm{IWB}}$.
Fig. \ref{ana1d2}  (b) shows a near breaking ion wave, 
with $E_{\rm{max}} \simeq E_{\rm{IWB}}$. 
In this case, at $\xi=0$, the ion density diverges, 
and the gradient of $\tilde{E}$ becomes infinite. 
It is essentially similar to the electron density and electric field 
in the case of electron wave breaking (see, e.g., Ref. \cite{dawson}).

\subsection{Ion trapping}

\subsubsection{1D PIC simulation}

\begin{figure}[htb]
\centering
\includegraphics[width=0.85\textwidth]{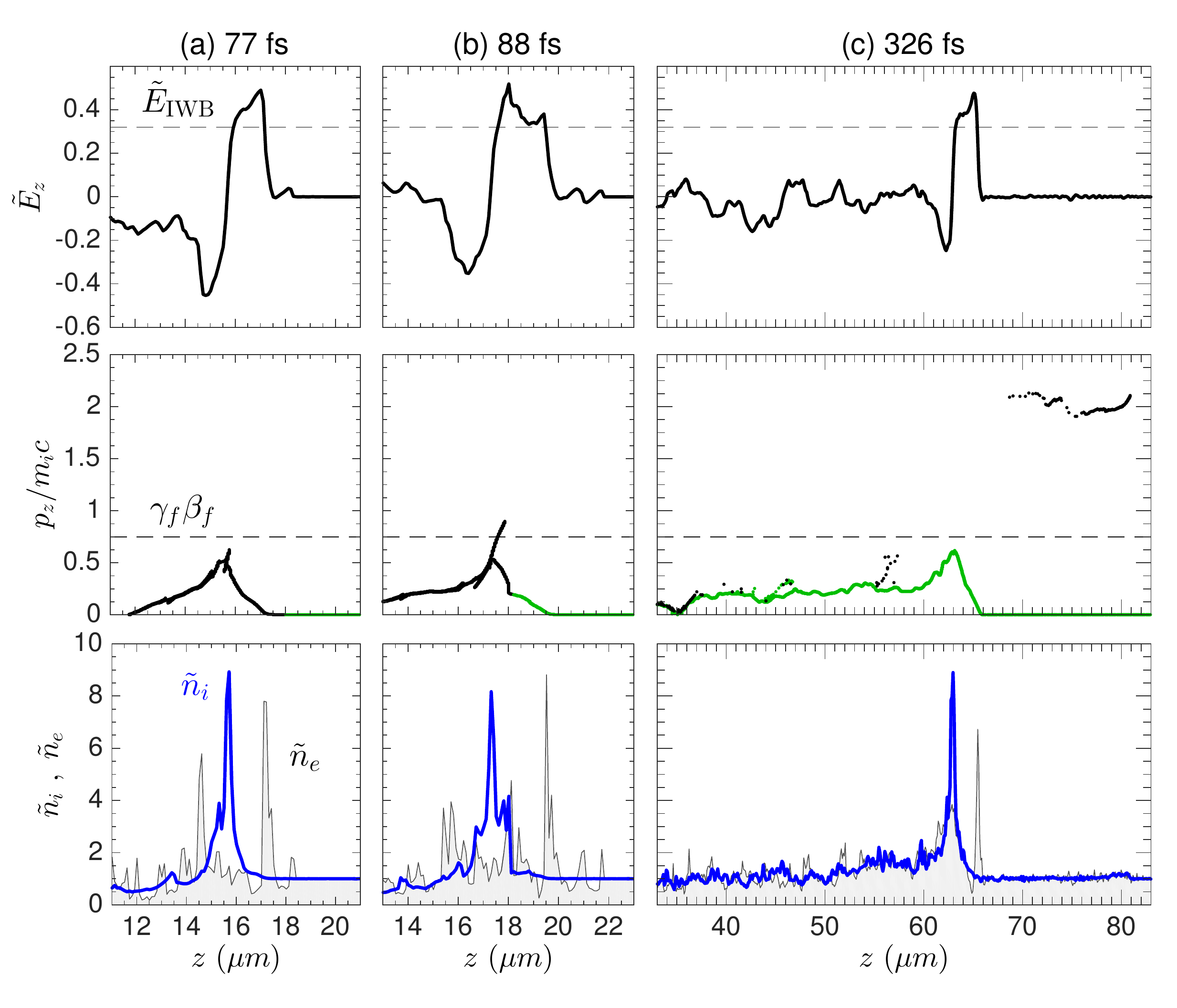}
\caption{
1D PIC simulation results 
of longitudinal electric field ($\tilde{E}_z$, top panels), 
ion momentum ($\gamma_i \beta_i$, middle panels), 
ion and electron densities ($\tilde{n}_i$, blue lines, 
and $\tilde{n}_e$, grey areas, lower panels), 
at different times (a) 77 fs, (b) 88 fs, (c) 326 fs, 
with the same laser pulse as that used in Fig. \ref{sim1d1} 
but a higher initial plasma density $n_0=3.2n_c$. 
There is no ion wave anymore and ion trapping happens. 
The horizontal dashed lines in top panels mark the 
threshold field for ion wave breaking $\tilde{E}_{\rm{IWB}} = 0.32$, 
which is clearly lower than the maximum electric field. 
In the middle panels, ions initially in the range of 
$z<18\mu m$ are plotted in black (dark) color, while other ions 
(initially $z>18\mu m$) are plotted in green (grey). 
The horizontal dashed lines in the middle panels 
mark the value of $\gamma_f \beta_f$. 
See text for more details. 
\label{sim1d2}
}
\end{figure}

In order to investigate ion trapping and acceleration 
under the condition of $E_{\rm{max}}>E_{\rm{IWB}}$, 
a 1D PIC simulation has been carried out  
with laser plasma parameters 
the same as that used in Fig. \ref{sim1d1} but
a higher initial plasma density $n_0=3.2n_c$. 
The simulation results are shown in Fig. \ref{sim1d2}. 
At 77 fs (Fig. \ref{sim1d2} (a)), 
a high density electron layer (in the region $17\mu m<z<17.5\mu m$) 
is produced at the front of the laser pulse (the laser pulse is not shown).  
The ion density peaks at about $15.7 \mu m$. 
The longitudinal electric field is positive in between 
the leading electron layer and the ion density peak  
and negative in the region behind. 
The maximum ion velocity is very close to 
the laser front velocity ($\beta_f \sim 0.6$). 
Ion trapping happens around the later time 88 fs (Fig. \ref{sim1d2} (b)). 
The average electron density 
in the range of $15\mu m<z<19\mu m$ in Fig. \ref{sim1d2} (b) lower panel 
is about $\sim 2n_0$.  
If we apply $\tilde{n}_{el} \sim 2$   
to Eq. (\ref{efthreshold}), 
we have $\tilde{E}_{\rm{IWB}} = 0.32$; 
it is marked in the top panels in Fig. \ref{sim1d2} by dashed lines. 
The maximum laser-driven charge-separation field 
is found to be $\tilde{E}_{\rm{max}} \sim 0.5$. 
It is clear that $\tilde{E}_{\rm{max}} > \tilde{E}_{\rm{IWB}}$. 
The trapped ions are accelerated by the charge-separation field, 
until they overtake the laser front (Fig. \ref{sim1d2} (c)).

\begin{figure}[htb]
\centering
\includegraphics[width=0.4\textwidth]{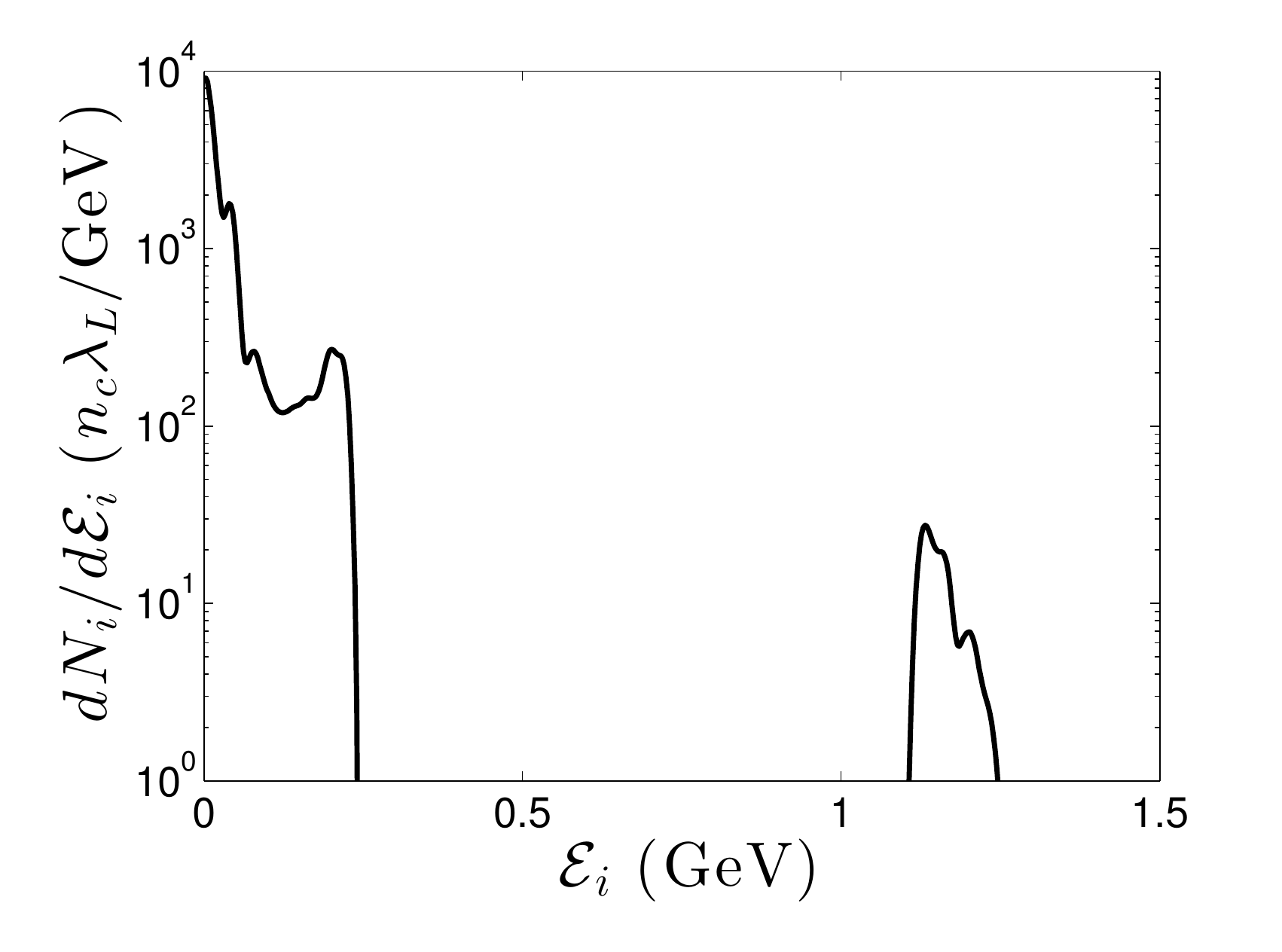}
\caption{Ion energy spectrum corresponding to the simulation result in 
Fig. \ref{sim1d2} (c). 
\label{spec1d2}
}
\end{figure}

In the frame comoving with the laser front  
ions come into the region behind of the laser front 
with initial velocity $-v_f$, then 
reflected and overtakes the laser front with velocity $v_f$. 
Therefore, in the laboratory frame, 
according to the relativistic velocity addition, 
we have the velocity of the output ions,   
$\beta_{i,\rm{out}} = 2 \beta_f/(1+\beta_f^2)$,  
and the kinetic energy \cite{iwba},
\begin{eqnarray}
\mathcal{E}_{i,\rm{out}} =  2 \gamma_f^2 \beta_f^2 m_i c^2. 
\end{eqnarray}
With $\beta_f \sim 0.6$, we have $\mathcal{E}_{i,\rm{out}} \sim 1.1$ GeV. 
This coincides with the peak of the energy spectrum directly observed 
from the simulation (Fig. \ref{spec1d2}).

Finally, the output ions are separated from the untrapped ions,
as is shown in Fig. \ref{sim1d2} (c) middle panel. 
It is noticed that all the output ions 
initially come from the region $z < 18 \mu m$. 
On the other hand, 
all the ions initially in the region $z > 18 \mu m$ 
have not been trapped and accelerated. 
Since the duration of the incident laser pulse is semi-infinite, 
this reflects that the ion trapping process is self-regulating and self-stops.  
This results in the peaked energy spectrum shown in Fig. \ref{spec1d2}.

\begin{figure}[htb]
\centering
\includegraphics[width=0.4\textwidth]{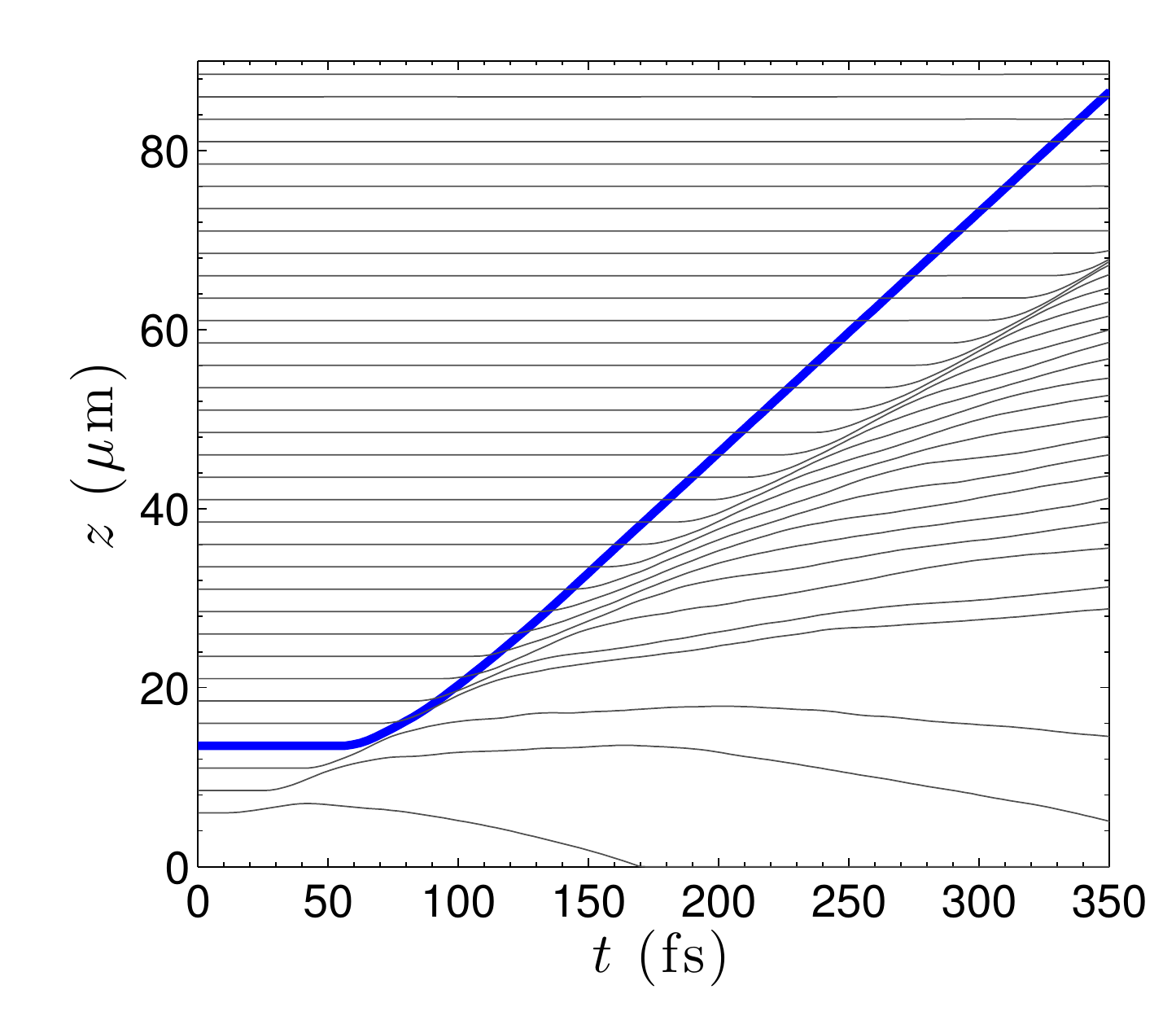}
\caption{Trajectories of ions with different initial positions 
for the 1D simulation shown in Fig. \ref{sim1d2}. 
The crossing trajectory with initial position $z(t=0)=13.5 \mu m$ is 
highlighted as a thick blue line. 
\label{traj1d2}
}
\end{figure}

Figure \ref{traj1d2} shows trajectories of ions. 
Different from that in Fig. \ref{traj1d1},  
trajectory crossing happens from $t\sim$ 80 fs onward. 
The trajectory of an ion initially 
at $z=13.5 \mu m$ (thick blue line) crosses with the trajectories 
of the ions initially at $z \geq 16 \mu m$ one by one. 
However, the trajectories of untrapped ions (thin black lines) do not 
cross each other,   
indicating that the untrapped ions 
still behave like a fluid.

\begin{figure}[htb]
\centering
\includegraphics[width=0.4\textwidth]{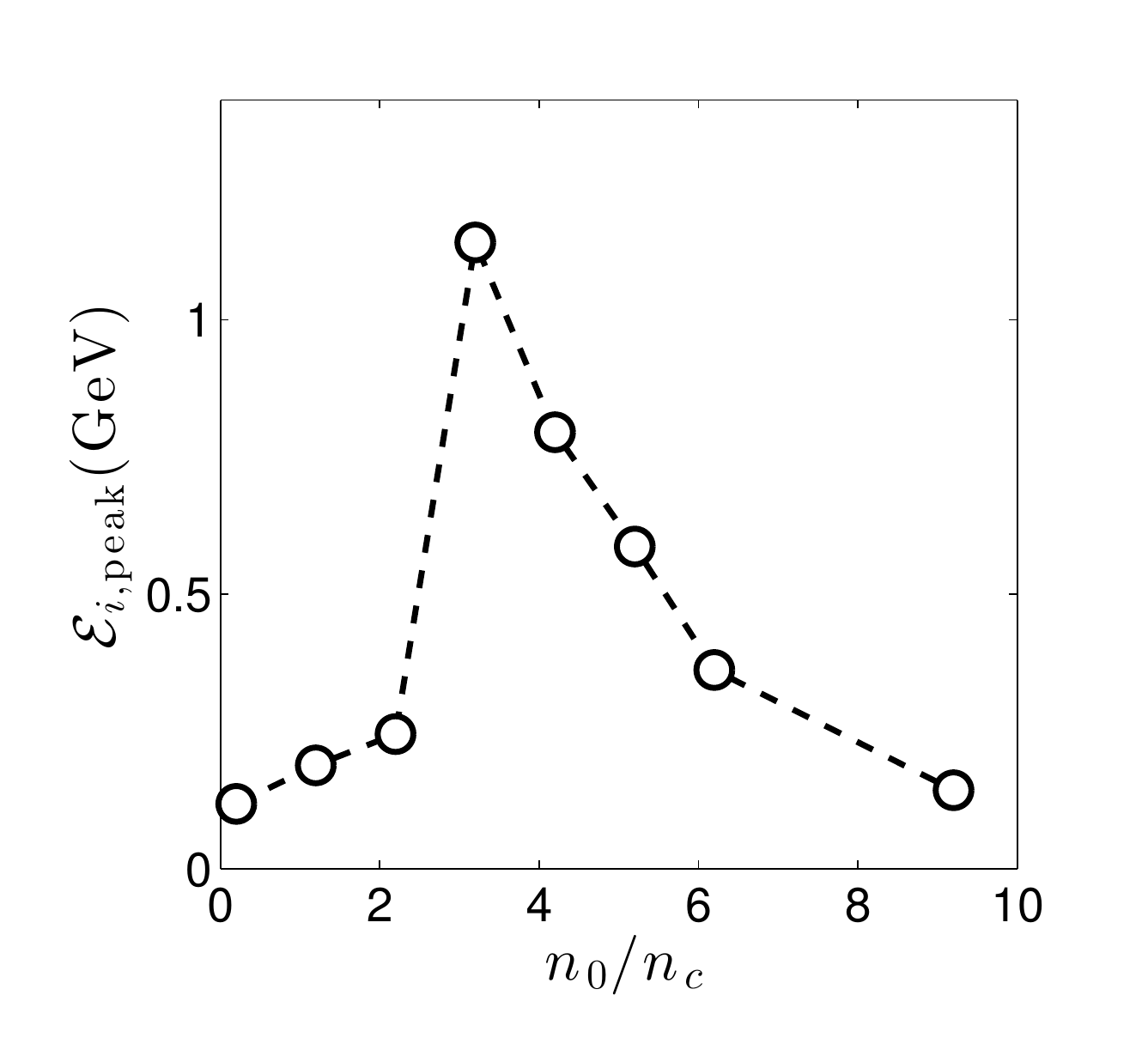}
\caption{
1D PIC simulation results of peak ion energy ($\mathcal{E}_{i,\rm{peak}}$)
with the same laser pulse ($a_0=44$)
as that used in Fig. \ref{sim1d1} and Fig. \ref{sim1d2} 
but different initial plasma densities $n_0$. 
\label{scan1d}
}
\end{figure}

Peak ion energies for different initial plasma densities are shown 
in Fig. \ref{scan1d}. 
For $n_0 \leq 2.2n_c$, 
there is no ion acceleration. They are in the regime of $E_{\rm{max}}<E_{\rm{IWB}}$. 
The maximum ion velocity is far less than the
corresponding laser front velocity.  
For $n_0 \geq 3.2n_c$, ion trapping happens and ions are 
accelerated. 
The peak energy of the output ion beam 
decreases with the increase of $n_0$.

\subsubsection{Self-stopping of ion trapping}

It is clear that the ion trapping discussed above (Figs. \ref{sim1d2}-\ref{scan1d}) happens in relativistically transparent regime. 
As is well known \cite{vrt1}, 
the laser front velocity in a relativistically transparent plasma 
is significantly faster than the hole-boring velocity $v_{\rm{hb}}$. 
This requires that the ion trapping process 
in relativistically transparent regime  has to self-stop. 
This can be explained by introducing a paradox. 
For a laser pulse with a long duration ($\tau_L\gg (\xi_a-\xi_c)/\omega_i$),  
if the self-trapping process is continuous from the beginning to the end, 
the total number of the accelerated ions can be estimated as 
$\mathcal{N}_i \sim n_0 v_f \Delta t $, 
where $\Delta t \sim \tau_L/(1-\beta_f)$ is the interaction time.
Then the total ion energy satisfies 
\begin{eqnarray}
\mathcal{N}_i \mathcal{E}_{i,\rm{out}}
&=& 2 n_0  m_i c^2 
\gamma_{f}^2 \beta_{f}^2  v_f  \Delta t
> 2 n_0  m_i c^2 
\gamma_{\rm{hb}}^2 \beta_{\rm{hb}}^2  v_f  \Delta t  \notag    \\
&=& \frac{2\beta_f(1-\beta_{\rm{hb}}) I_0 \tau_L }{(1-\beta_f)(1+\beta_{\rm{hb}})}  
> \frac{2\beta_f}{1+\beta_f} I_0 \tau_L,  
\label{paradox}
\end{eqnarray}
where we have used the hole-boring velocity 
$\beta_{\rm{hb}} = v_{\rm{hb}}/c = 1/(1+\sqrt{\mu n_0/n_c}/a_0)$ \cite{robinsonhb}.  
Relation (\ref{paradox}) is impossible because 
the term on its right hand side 
is the total energy transferred from 
the laser pulse to the plasma particles via laser radiation pressure 
\cite{radpre} 
and cannot be less than the total ion energy. 
Therefore,
the ion trapping has to self-stop in a short time period
($\Delta t < \tau_L/(1-\beta_f)$).  
This has been observed
in simulations in several separate works \cite{weng,robinson,iwba,sasl3} 
and makes ion acceleration in relativistically transparent plasma different 
from that in the hole-boring regime \cite{footnote2}.

Intrinsically, the reason for the self-stopping of ion trapping is that 
the electric field produced by the trapped ions reduces the 
acceleration of subsequent ions, 
so that their maximum velocity 
is less than the laser front velocity
and they cannot be trapped anymore. 
This is similar to the beam-loading effect in 
electron wake-field acceleration \cite{ewbrev}.

\subsubsection{Model of ion trapping}

\begin{figure}[htb]
\centering
\includegraphics[width=0.45\textwidth]{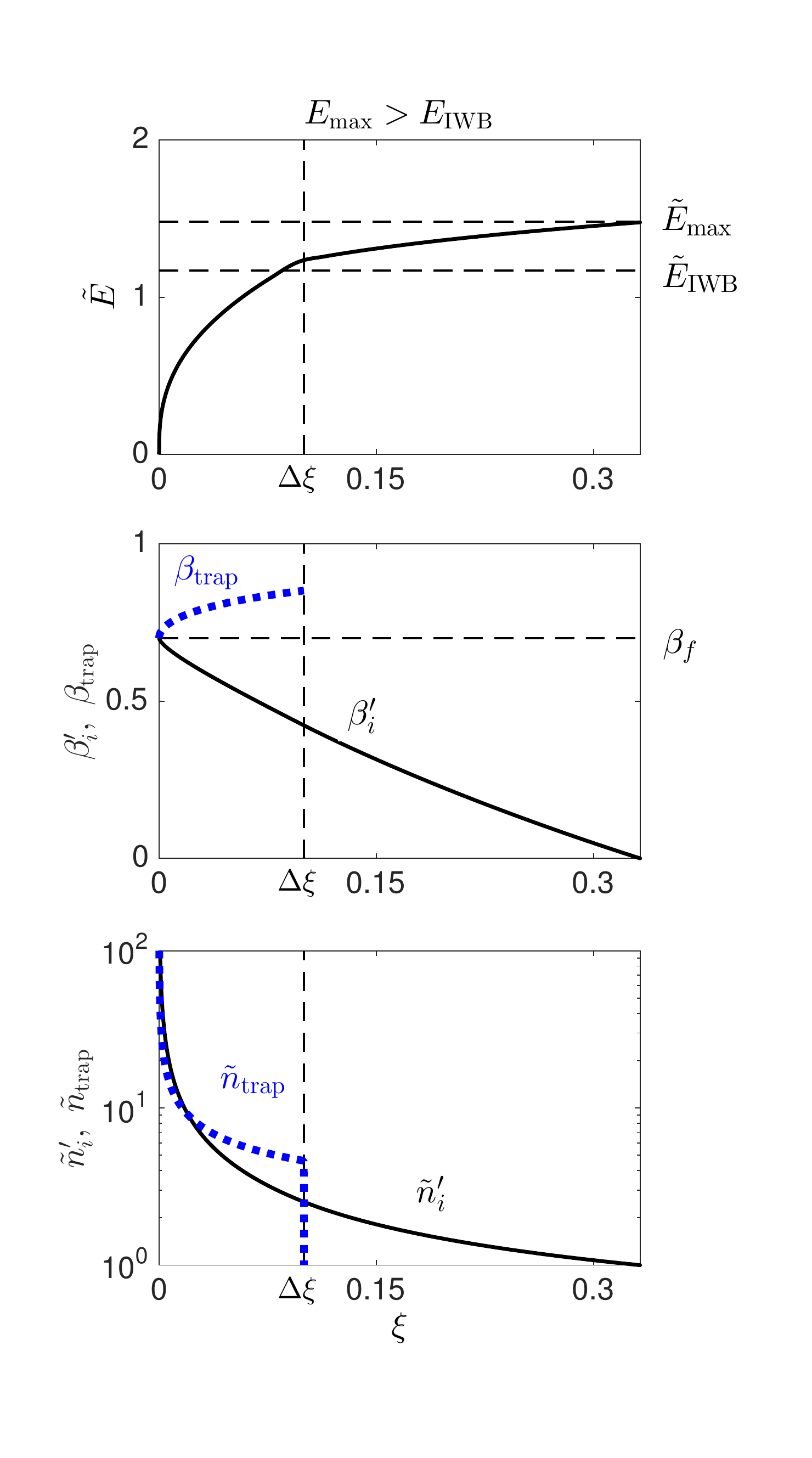}
\caption{ 
Results of the model of 
(top panel) the longitudinal electric field $\tilde{E}$, 
(middle panel) the velocity and (lower panel) density 
of un-trapped ($\beta_i'$ and $n_i'$, black sold lines) 
and trapped ($\beta_{\rm{trap}}$ and $\tilde{n}_{\rm{trap}}$, 
thick blue dotted lines) ions  
when ion trapping just stopped  
($\beta_i(0)-\beta_f \to 0^-$ and $\beta_{\rm{trap}}(0)-\beta_f \to 0^+$).  
The results are obtained 
by numerically solving Eqs. (\ref{eqbetai2}-\ref{ntrapdef})
with $\Delta \xi =0.1$ and the parameters 
$\beta_f$, $n_{e}$, $v_b$ the same as those used in Fig. \ref{ana1d2}.  
The results are under the condition of $E_{\rm{max}}>E_{\rm{IWB}}$. 
The upper horizontal dashed line in the top panel marks the maximum electric field
$\tilde{E}_{\rm{max}}=1.5$, while the lower one marks 
$\tilde{E}_{\rm{IWB}}=1.2$. 
The vertical dashed lines are located at $\xi = \Delta \xi =0.1$. 
\label{ana1d3}
}
\end{figure}

When ion trapping is finished, the untrapped ions 
still behave like a fluid (see Fig. \ref{traj1d2}). 
We focus on the conditions that ion trapping was just finished
and the size of the trapped ion beam is small 
so that the relative motion between the trapped beam 
and the charge-separation field 
is very slow compared with the laser front 
propagation in the laboratory frame. 
Therefore the untrapped ions 
and the charge-separation field are still in quasi-static state 
in a short time period.  
We denote the velocity and density of the untrapped ions 
by $v_i'$ and $n_i'$, and those of the the trapped ions by 
$v_{\rm{trap}}$ and $n_{\rm{trap}}$, respectively. 
The untrapped ions satisfy the fluid equations of motion and continuity, 
which are then written as
\begin{eqnarray}
(1-\beta_i'/\beta_{f})\frac{d \gamma_i' \beta_i'}{d\xi}&=&  -\tilde{E},
\label{eqbetai2}  \\
\tilde{n}_i' &=& \frac{1}{1-\beta_i'/\beta_{f}}, 
\label{eqni2}
\end{eqnarray}
while Poisson’s equation now includes the trapped charge, 
\begin{eqnarray}
\frac{d \tilde{E}}{d\xi} =  \beta_{ f  }  
 (\tilde{n}_i' + \tilde{n}_{\rm{trap}} - \tilde{n}_e ). 
\label{eqeztr}
\end{eqnarray}
The trapped ions move in the charge-separation field according to 
the equation of motion,
\begin{eqnarray}
(1-\beta_{\rm{trap}}/\beta_{ f  }) 
\frac{d \gamma_{\rm{trap}} \beta_{\rm{trap}} }{d\xi }=  -\tilde{E}. 
\label{eqbtrap}
\end{eqnarray}
In the frame comoving with the laser front, 
self-injection of ions can be seen as a process of reflection. 
Charge conservation then gives \cite{footnote3}
$\tilde{n}_i'(z) (\beta_i'(z)-\beta_f)
+\tilde{n}_{\rm{trap}}(z) (\beta_{\rm{trap}}(z) -\beta_f)=0$
for $z$ in the region where $\tilde{n}_{\rm{trap}}(z)\neq 0$. 
It holds until the ion trapping self-stops and Eq. (\ref{eqni2}) applies. 
Therefore, we get the density of trapped ions when the ion trapping just stopped, 
\begin{eqnarray}
n_{\rm{trap}} = 
\begin{cases}
  1/(\beta_{\rm{trap}}/\beta_f -1), & 0\leq \xi \leq  \Delta \xi; \\
  0, & \xi<0 ~ {\rm{or}} ~ \xi>\Delta \xi, 
\end{cases}
\label{ntrapdef}
\end{eqnarray}
where $\Delta \xi$ denotes the length of the trapped ion beam. 
Fig. \ref{ana1d3} shows the electric field $\tilde{E}$, 
the ion velocity ($\beta_i'$ and $\beta_{\rm{trap}}$) 
and density ($\tilde{n}_i'$ and $\tilde{n}_{\rm{trap}}$)
by numerically solving 
Eqs. (\ref{eqbetai2}-\ref{ntrapdef}). 
Due to the trapped charge, 
the electric field in the region $0 \leq \xi \leq \Delta \xi$ 
is significantly increased in comparison to  that in 
Fig. \ref{ana1d2} (b).
The increase in the electric field eventually closes the gap 
between the maximum field $E_{\rm{max}}$
driven by the laser pulse and 
the intrinsic wave breaking field $E_{\rm{IWB}}$.

According to Eqs. (\ref{eqbetai2},\ref{eqni2},\ref{eqeztr}), 
the analogue to Eq. (\ref{int1}) now reads
\begin{equation}
\frac{d}{d \xi}\left(\frac{ \tilde{E}^2}{2}
+  \beta_{ f  } \gamma_i' \beta_i' 
+ \tilde{n}_e \gamma_i' (1 - \beta_f \beta_i' )
\right) =\beta_{ f  } \tilde{E} \tilde{n}_{\rm{trap}}.   
\label{int2}
\end{equation}
Since ion trapping is finished, there is no further self-injection of ions, i.e.,
the maximum velocity of the untrapped ions must be smaller than  or equal to the laser front velocity ($\beta_i'(0) \leq \beta_f$). 
Integrating Eq. (\ref{int2}) from $\xi=0$
to $\xi=\xi_b$, 
and making use of Eq. (\ref{efthreshold}), 
we get
\begin{eqnarray}
\beta_f\int_0^{\xi_b} \tilde{n}_{\rm{trap}}(\xi) \tilde{E} d\xi 
\geq \tilde{E}_{\rm{IWB}}^2 \chi /2, 
\label{eftrap00}
\end{eqnarray}
where 
\begin{eqnarray}
\chi=E_{\rm{max}}^2/E_{\rm{IWB}}^2-1  
\label{chi}
\end{eqnarray}
characterizes the strength of ion wave breaking. 
Relation (\ref{eftrap00}) is the condition of no ion injection. 
It is a generalization of the condition of no ion trapping 
$\tilde{E}_{\rm{max}} \leq \tilde{E}_{\rm{IWB}}$. 
When ion trapping just self-stopped, the amount of the trapped ions is 
the minimum to make relation (\ref{eftrap00}) valid, thus    
the equal sign has to be taken, 
\begin{eqnarray}
\beta_f\int_0^{\Delta \xi} \tilde{n}_{\rm{trap}}(\xi) \tilde{E} d\xi 
= \tilde{E}_{\rm{IWB}}^2 \chi /2.  
\label{eftrap0}
\end{eqnarray}
Equation (\ref{eftrap0}) connects the trapped ions with 
the laser-driven maximum electric field.

\section{Ion trapping in 3D geometry}

In practical 3D geometry, 
complexity arises due to transverse effects such as 
laser self-focusing and electron evacuation. 
However, ion trapping is essentially an 1D wave breaking process.
It occurs localized close to the laser axis 
(see 3D simulations in Ref. \cite{iwba} and Fig. \ref{sim3d}), 
especially when the strength of ion wave breaking is small ($\chi \ll 1$). 
This is different from electron trapping in the bubble regime, 
where electrons are injected sideways and it is 
qualitatively different from that in 1D \cite{bubble1d3d}. 
We assume \cite{footnote4} that the laser spot size is large enough 
so that in its focus both charge-separation field and ion flow 
are mainly along the longitudinal direction 
($\nabla \cdot \boldsymbol{E} \sim \partial E_z/\partial z $
and $\nabla \cdot (n_i \boldsymbol{v}_i) \sim \partial n_i v_{i}/\partial z $).
Then the 1D model is still capable of describing trapping and acceleration of ions.  
The difference here is that electrons are expelled transversely 
by the laser pulse such that $n_{el} \ll n_i$. 
This makes the ion trapping problem solvable.

We consider the condition that ion trapping just stopped so that 
$\tilde{n}_i'$ is still singular at $\xi=0$. 
This allows us to try the power-law ansatz in the range of $0\leq \xi \leq \Delta$, 
\begin{equation}
\tilde{n}_i'(\xi) = \Lambda \xi^{-\Theta}, 
\label{ans1}
\end{equation}
with $\Lambda>0$ and $\Theta>0$. 
Then according to Eq. (\ref{eqni2}) and $\beta_i'(\xi) \leq \beta_f$, we have 
\begin{eqnarray}
1- \beta_i'(\xi)/\beta_f = \Lambda^{-1} \xi ^{\Theta}. 
\end{eqnarray}
Applying it to Eq. (\ref{eqbetai2}), and making use of the identity 
$d(\beta \gamma) = \gamma^3 d\beta$, we find
\begin{eqnarray}
\tilde{E}(\xi) = \beta_f \Theta \Lambda^{-2}\xi^{2\Theta -1} \gamma_i'^3. 
\end{eqnarray}
The Taylor series of $\gamma_i'$ at $\beta_i'=\beta_f$ is
\begin{eqnarray}
\gamma_i' = \gamma_f + \gamma_f^3 \beta_f (\beta_i'-\beta_f) +... . 
\end{eqnarray}
For simplicity we take the first-order approximation, 
which works very well when 
\begin{eqnarray}
 \gamma_f^3 \beta_f (\beta_f - \beta_i')  \ll \gamma_f. 
 \label{taylor1}
\end{eqnarray}
Hence, we have    
\begin{eqnarray}
\tilde{E}(\xi) = \beta_f \gamma_f^3  \Theta \Lambda^{-2}\xi^{2\Theta -1} . 
\label{ans3}
\end{eqnarray}
Similarly, when 
\begin{eqnarray}
\gamma_f^3 \beta_f (\beta_{\rm{trap}}-\beta_f) \ll \gamma_f,  
\label{taylor2}
\end{eqnarray}
we obtain the velocity of the trapped ions by 
taking Eq. (\ref{ans3}) into Eq. (\ref{eqbtrap}),   
\begin{eqnarray}
\beta_{\rm{trap}}(\xi)/\beta_f - 1 = \Lambda^{-1} \xi ^{\Theta}.  
\end{eqnarray}
Thus, according to Eq. (\ref{ntrapdef}), we have 
\begin{equation}
\tilde{n}_{\rm{trap}}(\xi) = \Lambda \xi^{-\Theta}. 
\end{equation}
Then Eq. (\ref{eqeztr}) becomes 
\begin{equation}
- \frac{\gamma_{ f  } ^3 \beta_{ f  } \Theta }{\xi^{1-\Theta} \Lambda}=
- \frac{\beta_{ f  } \xi^{1- 2\Theta} \Lambda^2}{(1-\Theta) }, 
\end{equation}
where the electron density has been ignored since  
$\tilde{n}_{el}\ll \tilde{n}_i$ in 3D geometry.  
Comparing the exponents of $\xi$ and front coefficients on both sides, 
we get
\begin{equation}
\Theta=2/3, 
\quad  \Lambda = 3^{-2/3} \gamma_{ f  }. 
\end{equation}
Finally, we have
\begin{eqnarray}
\tilde{E}(\xi) &=& 2   \beta_{ f  } \gamma_{ f  }(3 \xi)^{1/3}, 
 \label{ezrscale}    \\
(\beta_{\rm{trap}}(\xi)/\beta_f -1) = 
(1-\beta_i'(\xi)/\beta_{ f  })  &=& \gamma_{ f  }^{-1} (3\xi)^{2/3},
 \label{birscale}     \\
\tilde{n}_{\rm{trap}}(\xi) = 
\tilde{n}_i'(\xi) &=&   \gamma_{ f  }(3 \xi) ^{-2/3}, 
\label{nirscale}
\end{eqnarray}
and the total ion density 
\begin{eqnarray}
\tilde{n}_i(\xi) = \tilde{n}_{\rm{trap}}(\xi) + \tilde{n}_i'(\xi) = 
2 \gamma_{ f  }(3 \xi) ^{-2/3} . 
\label{ninrscale}
\end{eqnarray}
By taking Eq. (\ref{nirscale}) into Eqs. (\ref{taylor1}) and (\ref{taylor2}), 
we find that the scaling approximation  
(Eqs. (\ref{ezrscale}-\ref{ninrscale})) 
is valid under the condition of  
\begin{eqnarray}
\xi \ll \frac{1}{3 \gamma_f^{3/2}\beta_f^3}. 
\end{eqnarray}

The scaling result of $\tilde{n}_{\rm{trap}}$ in Eq. (\ref{nirscale}) 
helps us to calculate the length of the trapped ion beam 
according to Eq. (\ref{eftrap0}),
\begin{eqnarray}
\Delta \xi =\frac{1}{3} \left(\frac{\chi}{2}\right)^{3/2}
\left(\frac{\tilde{E}_{\rm{IWB}}}{\gamma_f\beta_f}\right)^3,      
\label{dxieq2}
\end{eqnarray}
and reduce the areal density of the trapped beam,  
\begin{equation}
\tilde{\mathcal{N}}_{\rm{trap}} 
=  \int_{0}^{\Delta \xi}\tilde{n}_{\rm{trap}}(\xi)d\xi 
=  (3\Delta\xi)^{1/3} \gamma_{ f  }.
\label{Nir}
\end{equation}
Then, in dimensional units with $\mathcal{N}_{\rm{trap}}
=(n_0v_f/\omega_i)\tilde{\mathcal{N}}_{\rm{trap}}$, we get 
\begin{equation}
q_i \mathcal{N}_{\rm{trap}}=\frac{q_i}{e} \epsilon_0E_{\rm{IWB}} 
\left(\frac{\chi}{2}\right)^{1/2}. 
\label{rhoi}
\end{equation}
Equation (\ref{rhoi}) is the central result of the present work. 
It gives the trapped charge per area  $q_i \mathcal{N}_{\rm{trap}}$ 
in terms of the charge number $q_i/e$,  
the displacement $\epsilon_0E_{\rm{IWB}}$, 
corresponding to the threshold electric field for ion wave breaking, 
and a factor dependent on the strength of ion wave breaking $\chi$.

Both $E_{\rm{max}}$ and $E_{\rm{IWB}}$
depend on the laser amplitude $a_0$ and the plasma density $n_0$.
For fixed $a_0$, 
the threshold for ion wave breaking is defined by
$E_{\rm{max}}(n_0^{\ast})=E_{\rm{IWB}}(n_0^{\ast})$
and allows to calculate the threshold density $n_0^{\ast}$. 
Here, we mark all quantities 
taken at the wave breaking threshold
by an asterisk.
Since $\chi(n_0^{\ast})=0$, 
by expanding $\chi(n_0)$ around $n_0=n_0^{\ast}$, we get
\begin{eqnarray}
\chi(n_0)=f_0 \left(\frac{n_0}{n_0^{\ast}}-1\right), 
\label{chiscale}
\end{eqnarray}
where $f_0= n_0^{\ast} \partial \chi(n_0^{\ast})/\partial n_0^{\ast}$. 
Then by using Eq. (\ref{efthreshold}) 
we find the number of the trapped ions per area in the form
\begin{equation}
\mathcal{N}_{\rm{trap}} \approx f_1 \frac{\epsilon_0E_{0}^{\ast}}{e} 
\left( \frac{n_0}{n_{0}^{\ast}}-1 \right)^{1/2},
\label{rhoifff}
\end{equation}
with 
$f_1= \beta_f^{\ast} \left( f_0 \gamma_f^{\ast} \left( 1 
- \delta^{\ast} \right) \right)^{1/2}$ and  
$\delta^{\ast}=\beta_b^{\ast} \gamma_b^{\ast} /\beta_f^{\ast}\gamma_f^{\ast}$, 
since $\tilde{n}_{el} = 0$. 
This result holds for fixed laser intensity and densities $n_0$ 
just beyond threshold density $n_0^{\ast}$. 
The precise value of the front factor $f_1$ in Eq. (\ref{rhoifff}) 
depends on the details not considered in the present paper. 
Comparing with the simulation results in Sec. \ref{seciv}, 
it turns out to be of order one. 
We emphasize that the scaling exponent $1/2$   
in Eq. (\ref{rhoifff}) traces back to the power-law exponents in 
Eqs. (\ref{ezrscale},\ref{nirscale}).

With the same approach,  
we also obtain the power-law scalings 
when ion wave breaking just sets in 
($E_{\rm{max}} \to E_{\rm{IWB}}$ and $\tilde{n}_{\rm{trap}}=0$), 
\begin{eqnarray}
\tilde{E}(\xi) &=& 2^{1/3}   \beta_{ f  } \gamma_{ f  }(3 \xi)^{1/3}, 
 \label{ezscale}    \\
 (1-\beta_i(\xi)/\beta_{ f  } ) &=& 2^{-1/3} \gamma_{ f  }^{-1}(3\xi)^{2/3},
 \label{biscale}     \\
\tilde{n}_i(\xi) &=& 2^{1/3} \gamma_{ f  }(3 \xi) ^{-2/3}.
\label{niscale}
\end{eqnarray}
One observes that the power-law exponents are the same as those 
in Eqs. (\ref{ezrscale},\ref{birscale},\ref{nirscale}), 
but the coefficients are different.
The change of the coefficients is a signal of ion trapping.

\section{3D PIC simulations}
\label{seciv}

\begin{figure}[htb]
\centering
\includegraphics[width=0.8\textwidth]{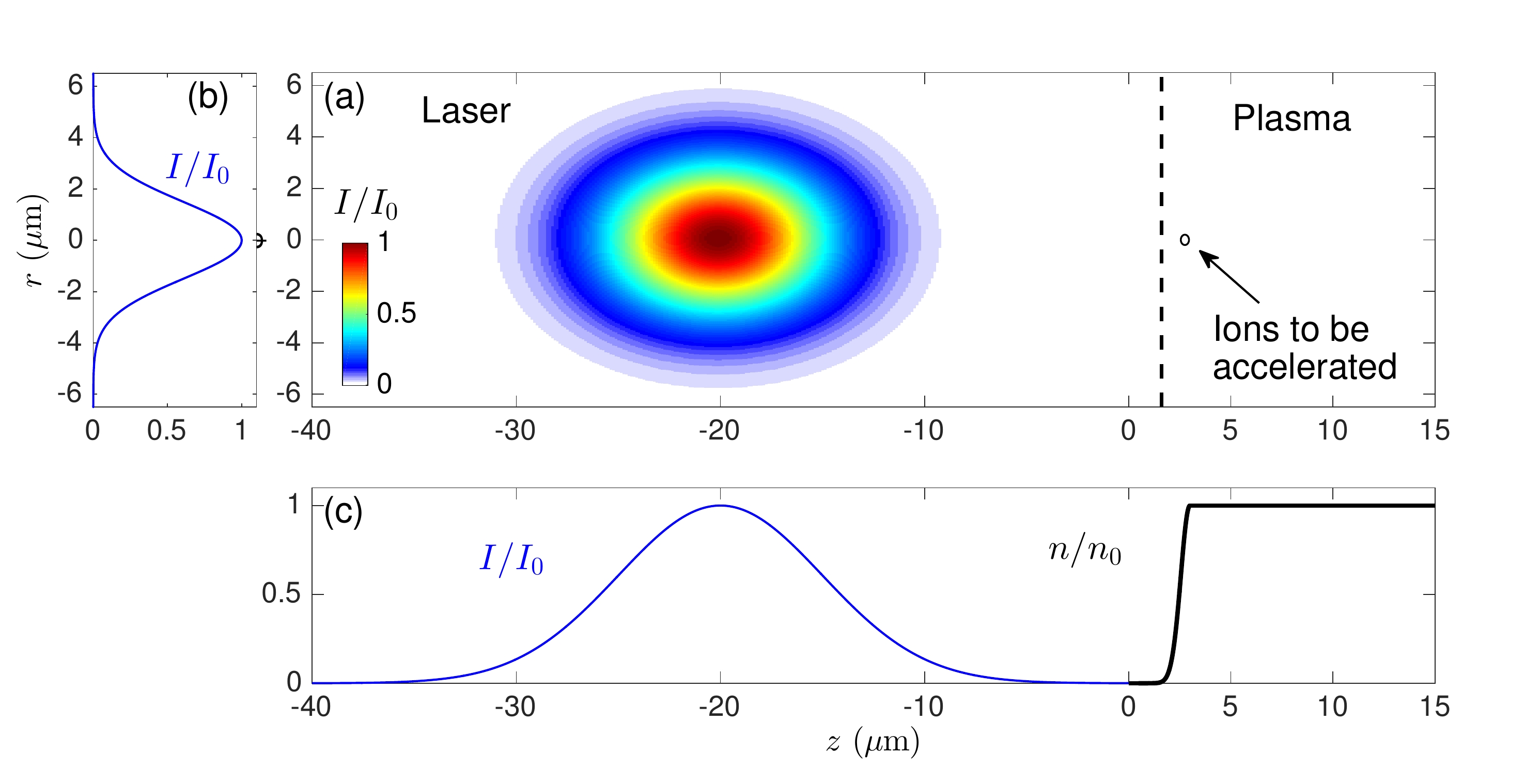}
\caption{
(a) Initial distributions of the laser intensity 
and the target. 
The laser pulse is 
circularly polarized with laser wave length $1\mu m$ and 
Gaussian intensity distribution 
in both (b) radial ($r$)  
and (c) longitudinal ($z$) directions. 
The intensity is  
$I=I_0\exp(-2r^2/R_L^2)\exp(-2(z-z_a)^2/(c\tau_L)^2)$, 
where $I_0=6\times 10^{20}\rm{W/cm^2}$, 
$R_L=3\mu m$, $z_a=-20\mu m$, and $\tau_L= 33.3$ fs. 
The plasma target consists of protons and electrons with  
plasma density $n_0=8\times 10^{21} {\rm{cm}}^{-3}$ 
uniformly distributed in the range of $z>3\mu m$ 
and a Gaussian profile density ramp  
$n=n_0 \exp(-(z-z_n)^2/\sigma_n^2)$ for $z<3\mu m$, 
where $z_n=3\mu m$ and
$\sigma_n=0.63\mu m$. 
Ions to be accelerated are initially in a small volume 
($2.7um<z<3um$ and $r<0.2 um$, see Fig. \ref{spec3d} and the corresponding text), 
as marked by the small black circle and the black arrow in (a).   
\label{para3d}
}
\end{figure}

We have carried out 
3D PIC simulations to verify the model 
predictions, using the plasma simulation code (PSC) \cite{psc}.
A circularly polarized laser pulse with wavelength
$\lambda_L=1\rm{\mu m}$ is vertically incident 
on hydrogen plasma  ($m_i/m_e=1836$ and $q_i/e=1$) 
with bulk density $n_0=7.2n_c$. 
At the surface, the density rises in the region $0<z<z_n$
according to $n(z)=n_0 \exp(-(z-z_n)^2/\sigma_n^2)$ 
with $z_n=3\rm{\mu m}$ and $\sigma_n = 0.63 \rm{\mu m}$. 
Such a slow rising boundary is reasonable in describing a 
target that is pre-heated by a laser pre-pulse. 
We have also run simulations with different $\sigma_n$ 
and found that  
results are insensitive to $\sigma_n$ 
for $0\leq \sigma_n \leq 1 \mu m$.
The peak laser intensity is 
$I_0=6\times 10^{20}\rm{W/cm^2}$, 
corresponding to a laser amplitude $a_0 =15$.
We have chosen a relatively low value of intensity
to demonstrate that ion trapping by ion wave breaking 
can already be studied with laser pulses 
presently available experimentally. 
Also, the shape of the pulse incident from the left side ($z=0$)
is modelled by a Gaussian amplitude in both radial direction 
$\left(r=\sqrt{x^2+y^2}\right)$  and time, 
$a(r,t)=a_0\exp(-r^2/R_L^2)\exp(-(t-t_0)^2/\tau_L^2)$
with $\tau_L =33.33$ fs (10 laser cycles),
$R_L=3 \rm{\mu m}$, and $t_0=66.67$ fs,
implying a moderate contrast ratio at the pulse front.
The initial distributions of laser intensity and plasma density
are shown in Fig. \ref{para3d}. 
In order to resolve the details of the ion motion, 
in particular near the wave breaking point,   
we have used 50 cells per micron and 
10 macro-particles per cell for each species.  
The initial temperatures are set to 10 keV for electrons
and 1 keV for ions.

\begin{figure}[htb]
\centering
\includegraphics[width=0.85\textwidth]{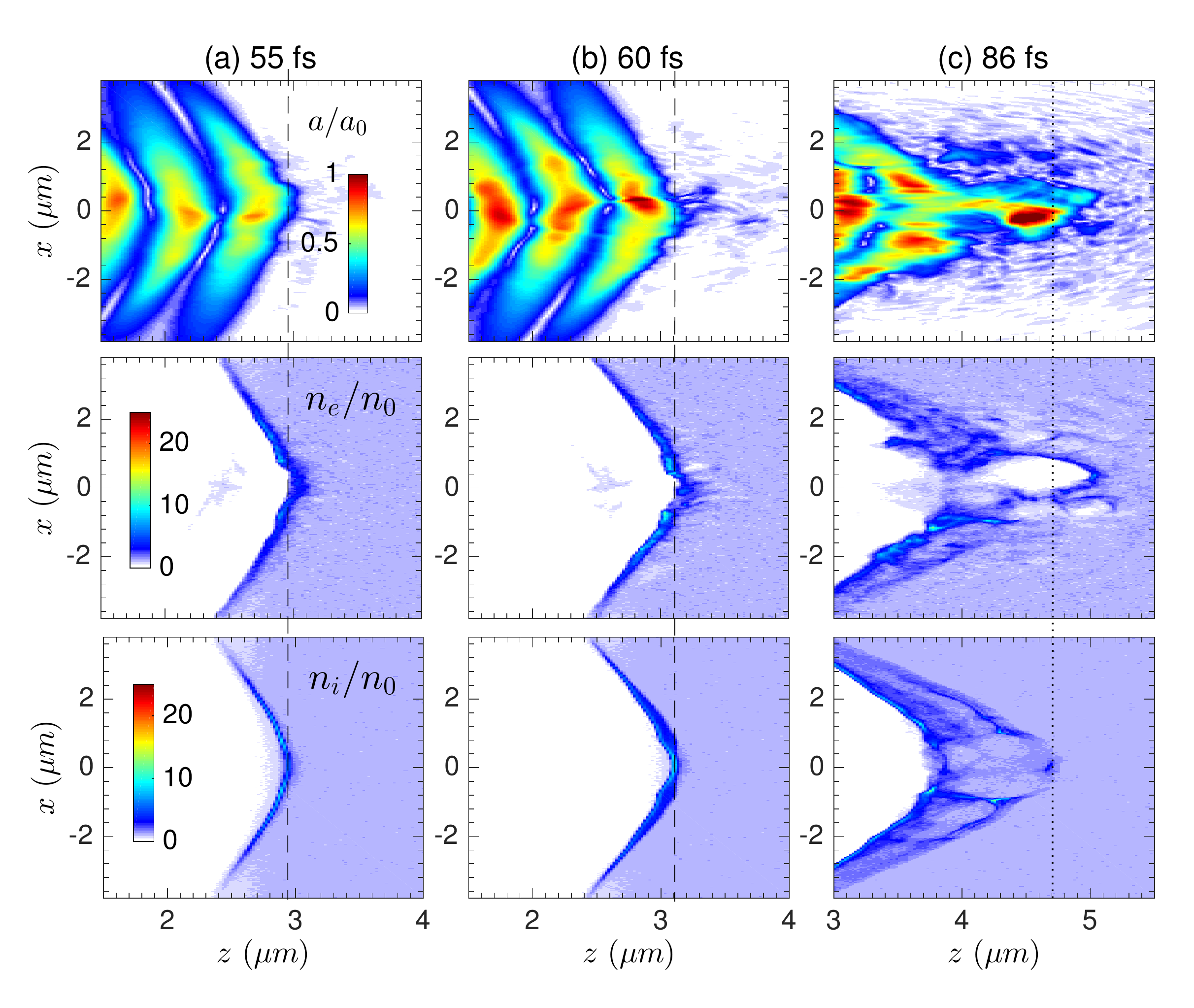}
\caption{
3D-PIC simulation results shown as cuts in $zx$-plane at $y=0$. 
From top to bottom,  
distributions of laser amplitude $a/a_0$, 
electron density $n_e/n_0$, 
and ion density $n_i/n_0$. 
Column (a) refers to the onset of ion trapping at 55 fs.
Column (b) shows the results at 60 fs. 
The dashed vertical lines in the columns (a) and (b)
mark the ion wave breaking points.
Column (c) shows the results at 86 fs,  
when trapping already stops and the trapped ion bunch 
marked by the dotted vertical line
is still accelerating. 
The initial profiles of laser intensity and plasma density are shown in Fig. \ref{para3d}.
\label{sim3d}
}
\end{figure}

\begin{figure}[htb]
\centering
\includegraphics[width=0.85\textwidth]{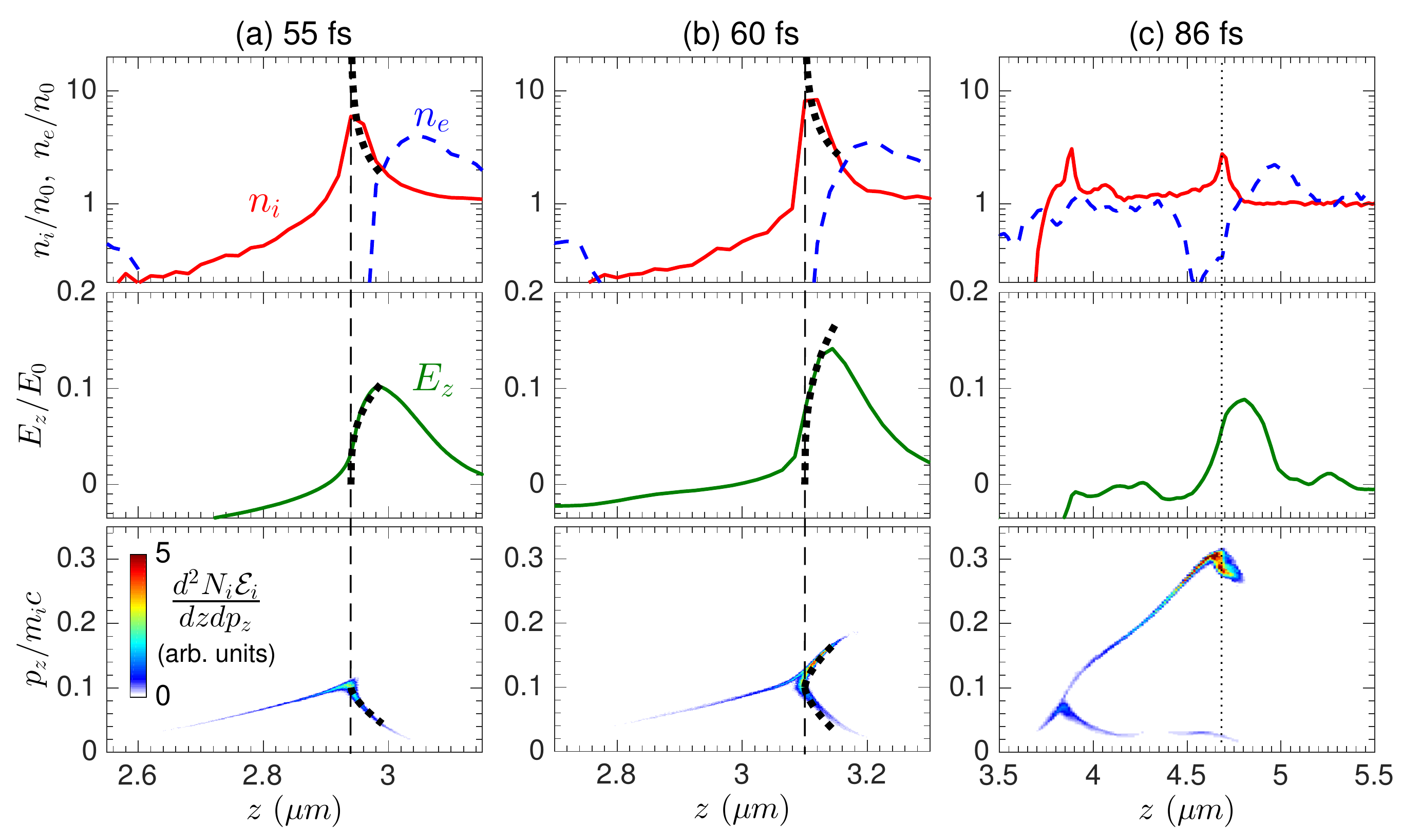}
\caption{
Comparison of on-axis ion evolution from   
the simulation in Fig. \ref{sim3d} 
with our model
(black dotted curves in the first two columns)
near the point of ion wave breaking, 
showing 
(top panels) ion density  $n_i/n_0$ (red solid lines), 
electron density $n_e/n_0$ (blue dashed lines), 
(middle panels) longitudinal electric field $E_z/E_0$ (green solid lines), 
and (lower panels) ion kinetic energy density
($d^2 N_i\mathcal{E}_i /dz d p_z$ in arbitrary units) 
in phase space $(z, p_z/m_ic)$. 
The simulation results are obtained by taking 
average  over $ r \leq 0.2\rm{\mu m}$. 
The dashed vertical lines in columns (a) and (b) mark 
the ion wave breaking points.
The dotted vertical line in column (c) marks 
the position of the trapped ion bunch.  
\label{axis3d}
}
\end{figure}

\begin{figure}[htb]
\centering
\includegraphics[width=0.45\textwidth]{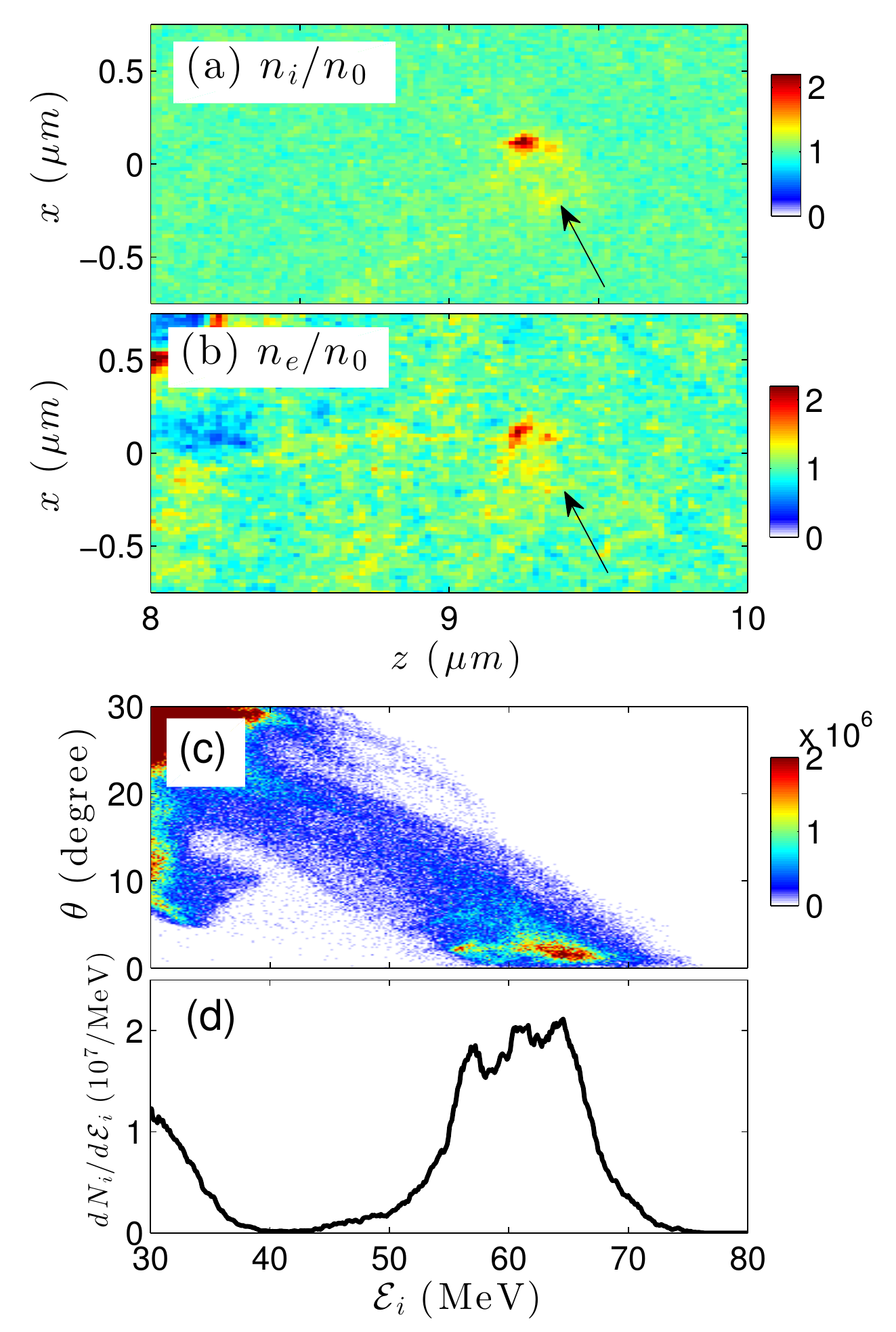}
\caption{Results of the output ion beam at 130 fs 
for the simulation in Fig. \ref{sim3d}.   
Distributions of (a) ion density and (b) electron density
as cuts in $zx$-plane at $y=0$.
(c) Ion angular-energy ($\theta$-$\mathcal{E}_i$) 
distribution ($d^2N_i/d\theta d\mathcal{E}_i$ in arbitrary units)
and (d) ion energy spectrum ($dN_i/d\mathcal{E}_i$) 
for ions within $\theta<10$ degrees. 
The ion beam and the corresponding electron beam 
in (a) and (b) are marked by 
black arrows, respectively. 
Most of the accelerated ions initially come from the small region  
marked by a black arrow in Fig. \ref{para3d} (a). 
\label{spec3d}
}
\end{figure}

The simulation results 
are shown in Figs. \ref{sim3d} and \ref{axis3d} 
as snapshots at different times.  
Figure \ref{sim3d} shows the distributions of laser field, 
electron density, and ion density in $zx$-plane, 
while Fig. \ref{axis3d} shows the density of electrons and ions, 
the longitudinal electric field, and 
the ion phase space ($p_z/m_ic,z$) 
averaged over $r\leq 0.2 ~\rm{\mu m}$ along the laser axis. 
At 55 fs (column (a)),
the front rising part of the laser pulse with relatively low-intensity 
has been depleted (Fig. \ref{sim3d} (a) top panel), and 
the laser front propagates forward with a relatively stable  
velocity $v_f \sim 0.1c$. It varies slowly until $t \sim 65 fs$. 
The ion wave develops a sharp peak in density
at $z\approx 2.94~\rm{\mu m}$ 
and is about to break.
The maximum ion velocity 
is almost equal to 
the laser front velocity (Fig. \ref{axis3d} (a) bottom panel).
The local distributions near the ion wave peak  
match well with the power-law scalings given by 
Eqs. (\ref{ezscale},\ref{biscale},\ref{niscale}) 
(dotted black lines in Fig. \ref{axis3d} (a)).
The charge-separation field increases with time. 
The strongest charge-separation field is observed
at 60 fs (column (b)), with the field maximum 
$E_{\rm{max}} \sim 0.14 E_0$. 
Ion trapping is now in progress. 
A good portion of the ions
has already been trapped 
and is accelerating in the region $z>3.1~\rm{\mu m}$.
Ion density and electric field have increased in agreement 
with the results of our model, now given 
by Eqs. (\ref{ezrscale},\ref{birscale},\ref{nirscale}). 
With the interaction time increasing  
the laser self-focusing effect becomes important.
It leads to an increase of laser intensity
on axis and enhances the acceleration of the trapped ions. 
Later at 86 fs (column (c)),
the focused laser pulse penetrates into the plasma and expels 
electrons, producing a cavity with relatively low electron density 
(Fig. \ref{sim3d} (c) middle panel). 
Ions dragged by the expelled electrons  
form a high density shell (Fig. \ref{sim3d} (c) bottom panel). 
An ion bunch is observed at the point $r \simeq 0$ and $z\simeq 4.7 \rm{\mu m}$, 
both in Fig. \ref{sim3d} (c) bottom panel and Fig. \ref{axis3d} (c) top 
and bottom panels.
The bunch propagates to the right with speed $\approx 0.3c$, 
faster than the laser front velocity, which now is  
$v_f \sim 0.2 c$ due to the laser self-focusing.
This corresponds to a proton energy of $(40\pm 10)$ MeV at this time. 
The ion beam will 
further gain energy (reaching about 65 MeV, see Fig. \ref{spec3d} (d)),
before overtaking the laser front and then freely cruising to the right.

Since the trapped ion beam is localized close to the laser axis 
and mainly accelerated forward, 
its areal density almost keeps constant during the acceleration process. 
At 86 fs, the trapped ion beam is already distinguished from background ions in both real space and phase space. 
As is seen in Fig. \ref{sim3d} (c) bottom panel and 
Fig. \ref{axis3d} (c) top and bottom panels,
the length of the ion beam is $\sim 0.2\mu m$, 
and the peak density is $\sim 8n_c$,  
corresponding to the ion number per area 
$e \mathcal{N}_{\rm{trap}} \sim 0.08 \epsilon_0 E_0$. 
This can be explained by the wave breaking model. 
According to the observations in Fig. \ref{axis3d} (b,c,e,f), 
we have  
$v_b\sim v_f/2 \sim 0.05c$  
(for example, in Fig. \ref{axis3d} (a), we chose $v_b$ as
the ion velocity at the point $z=3\mu m$ where $E_z$ peaks). 
By taking the parameters $a_0\sim 15$ and $n_0 \sim 7.2n_c$, 
and using Eq. (\ref{efthreshold}), we get 
$E_{\rm{IWB}} = 0.1 E_0$. 
Then according to Eq. (\ref{rhoi}) and making use of $E_{\rm{max}}\sim 0.14 E_0$ 
we have $e \mathcal{N}_{\rm{trap}} = 0.07 \epsilon_0 E_0$, 
close to the result directly observed in the simulation.

Ion acceleration stops when the trapped ion beam overtakes the laser front. 
After that, the output ion beam moves in the plasma freely,
accompanied by an almost equivalent electron beam, 
as shown in Fig. \ref{spec3d} (a,b). 
The neutralized beam can propagate in plasma
for a very long distance almost without energy loss. 
Furthermore, the ion beam is very compact. 
It is distributed in a small space volume ($\sim 0.2\mu m^3$)
with the peak density $\sim 8n_c$. 
It contains about $2.6 \times 10^8$ ions. 
The divergence angle is less than 
5 degrees (Fig. \ref{spec3d} (c)).  
The corresponding energy spectrum is quasi-mono-energetic and peaks
at about 65 MeV with an energy spread of about 10 MeV (Fig. \ref{spec3d} (d)). 
By tracking ions in the simulation we find that
most of the ions in the output ion beam initially come from 
a small spherical-like region ($2.7um<z<3um$ and $r<0.2 um$)  
marked by a black arrow in Fig. \ref{para3d} (a). 
This confirms that ion trapping mainly happens in 
the time period between 55 fs and 60 fs.

\begin{figure}[htb]
\centering
\includegraphics[width=0.4\textwidth]{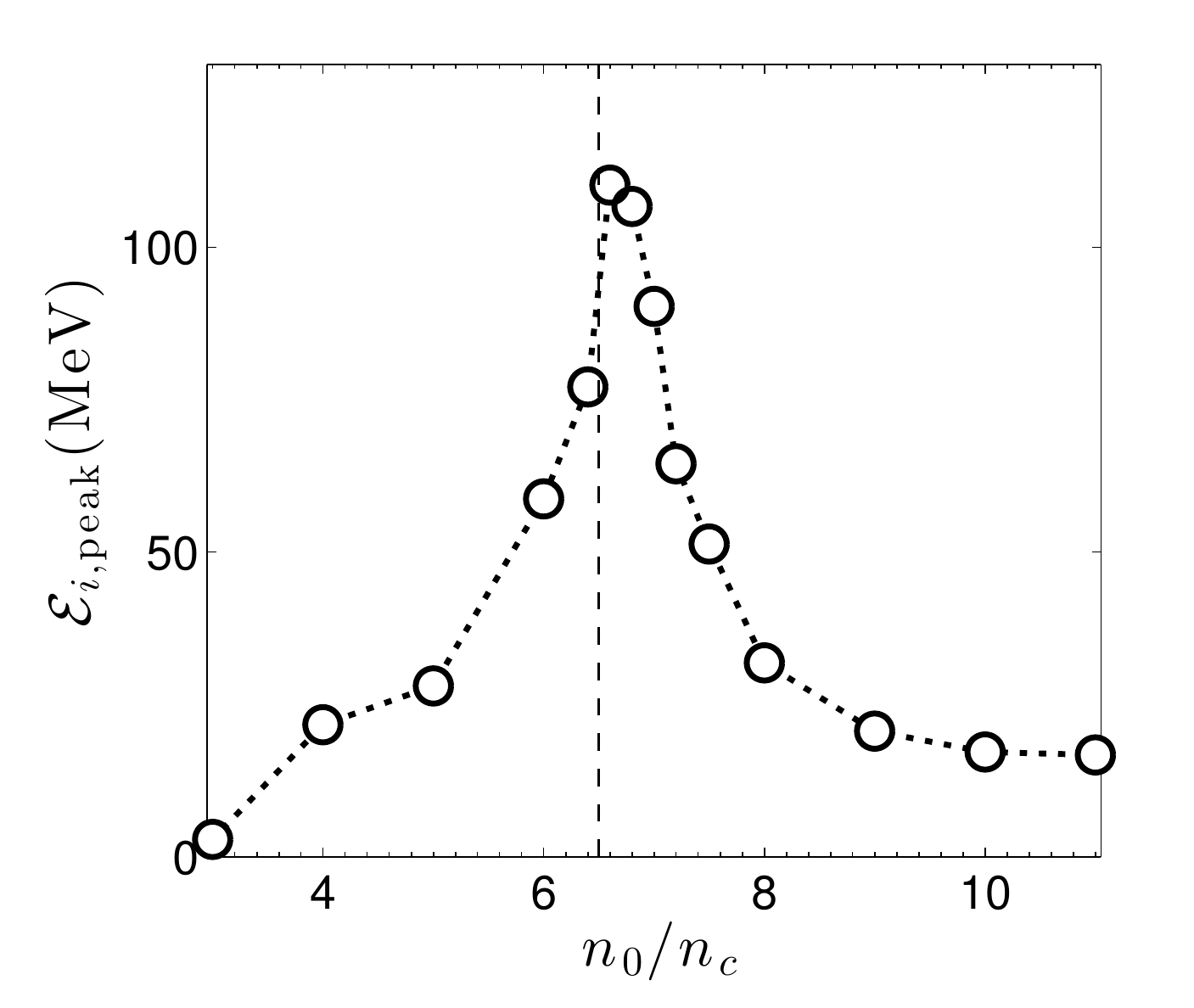}
\caption{3D PIC simulation results of  
Peak ion energy $\mathcal{E}_{i,\rm{peak}}$ (black circles), 
with the same laser pulse as that shown in Fig.  \ref{para3d},  
but different initial plasma densities $n_0/n_c$. 
The dashed vertical line at $n_0/n_c = 6.5$ 
marks the threshold density for ion trapping and acceleration. 
\label{scan3d}
}
\end{figure}

Simulation results of peak ion energy $\mathcal{E}_{i,\rm{peak}}$ 
for different initial plasma densities $n_0$
are shown in Fig. \ref{scan3d}.  
The most energetic ion acceleration occurs for 
initial plasma density $n_0=6.6n_c$, 
with peak energy about 110 MeV.  
For $n_0 \leq 6.4n_c$ there is no output ion beam  
on the laser axis and the final ion spectra have no clear peaks 
(the maximum ion energies are plotted instead). 
The threshold density for ion wave breaking is then obtained approximately as 
$n_0^{\ast}\sim 6.5n_c$, marked by a vertical dashed line in Fig. \ref{scan3d}. 
Beyond it, the peak ion energies decrease  
with the increase of the initial plasma density. 
For $n_0>8.5n_c$ the plasma becomes opaque and our model does not apply.

\begin{figure}[htb]
\centering
\includegraphics[width=0.4\textwidth]{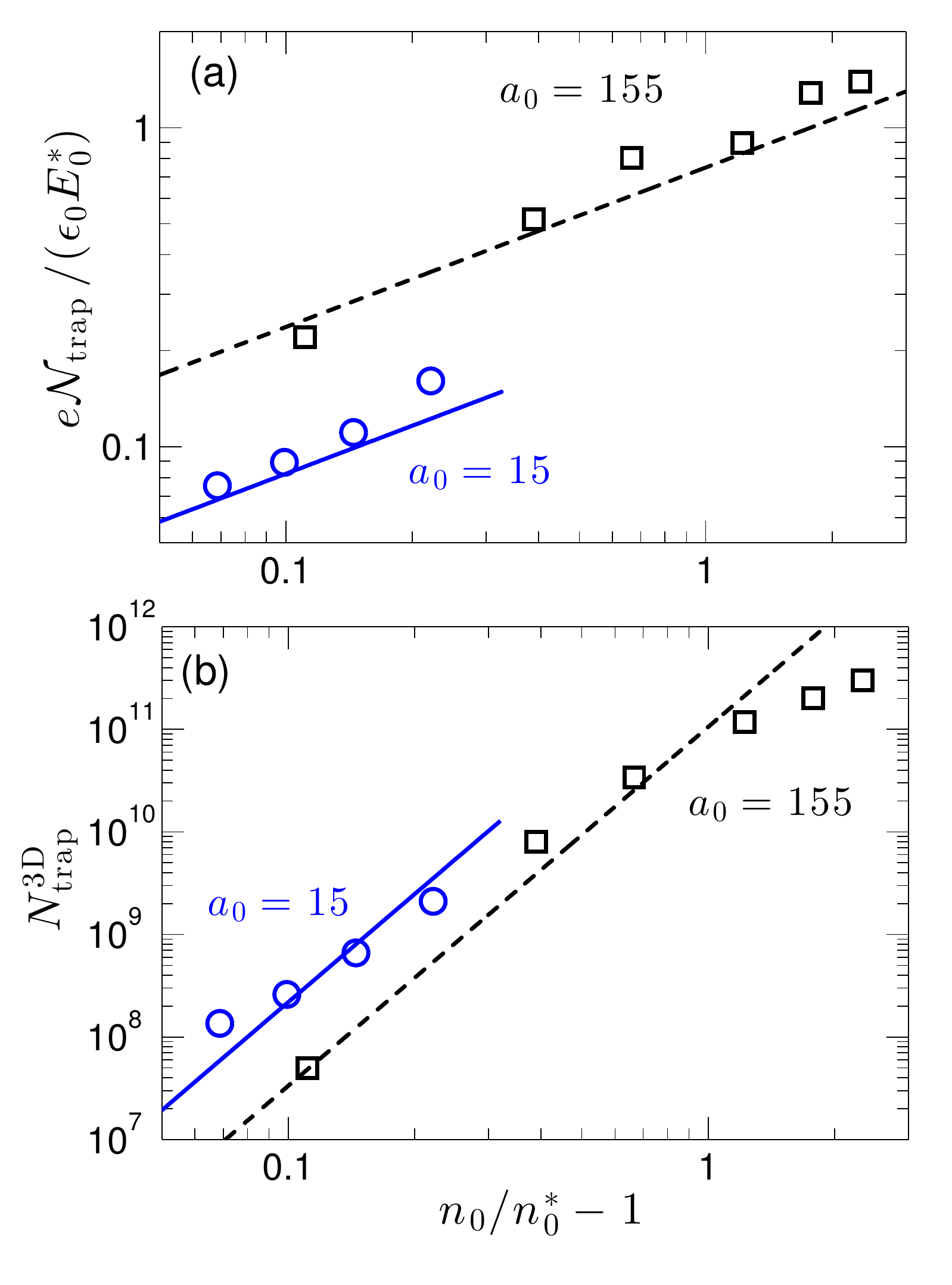}
\caption{
(a) The number of trapped ions per area  $\mathcal{N}_{\rm{trap}}$
and (b) the total number of the accelerated ions 
$N_{\rm{trap}}^{\rm{3D}}$
for fixed laser amplitude $a_0$ versus $(n_0/n_0^{\ast}-1)$,
where $n_0$ denotes the initial target density and $n_0^{\ast}$
the threshold density for ion trapping.
Circles and squares refer to 3D PIC simulations and the lines in (a)
to Eq. (\ref{rhoifff})  and in (b) to Eq. (\ref{nifff}); 
blue circles (case 1, $a_0=15$, $n_0^{\ast}=6.5n_c $) corresponding  
to the 3D simulations shown in Fig. \ref{scan3d} and 
black sqares (case 2, $a_0=155$, $n_0^{\ast}=1.8n_c$) 
to 3D simulations reported in Ref. \cite{iwba}.
\label{scale}
}
\end{figure}

In Fig. \ref{scale} (a), we compare values of $\mathcal{N}_{\rm{trap}}$
from simulations with those from Eq. (\ref{rhoifff}) 
for laser pulses having very different amplitudes: 
Case 1 with $a_0=15$ 
(blue, Gaussian distributions in both space and time, 
shown in Fig. \ref{para3d}) 
and case 2  with $a_0=155$ 
(black, Gaussian distribution in space and super-Gaussian in time, 
reported in Ref. \cite{iwba}). 
They correspond to threshold densities $n_0^{\ast} \simeq 6.5n_c$ for case 1 
and $n_0^{\ast} \simeq 1.8n_c$ (peak ion energy about 6 GeV) for case 2. 
The simulation results are well described by the square root scaling,
represented by the straight lines with slope 1/2 
in the double-logarithmic plot. 
The position of these lines has been adjusted 
by choosing $f_1=0.25$ for case 1 
and $f_1=0.75$ for case 2 in Eq. (\ref{rhoifff}).
It demonstrates that the present scaling result
holds over laser intensities ranging from 
$10^{20}$ to $10^{22}$W/cm$^2$.

The 3D PIC simulations provide us 
with absolute numbers of accelerated ions; 
they are plotted in Fig. \ref{scale} (b). 
In simulations we have found that for a given laser pulse 
the trapped ion beams  
appear as bullet-like bunches 
with transverse sizes $\Delta r$ 
scale roughly linearly with
the longitudinal ones $\Delta z$, 
especially when $\chi \ll 1$.   
The longitudinal size is estimated as
$\Delta z=(c\beta_f^{\ast}/\omega_i^{\ast})\Delta \xi$, 
where   
$\Delta \xi = 
\left(f_0 \left(1- \delta^{\ast}  \right) 
\left(n_0/n_0^{\ast}-1\right)/\gamma_f^{\ast}\right)^{3/2}/3$, 
obtained by applying Eqs. (\ref{efthreshold}) and (\ref{chiscale}) 
to Eq. (\ref{dxieq2}). 
Therefore, we estimate the absolute number 
of trapped and accelerated ions as
\begin{eqnarray}
 N_{\rm{trap}}^{\rm{3D}} 
\approx (\Delta r)^2\mathcal{N}_{\rm{trap}}
\approx f_2 
\left( \frac{c}{\omega_i^{\ast}} \right)^2 \frac{\epsilon_0 E_0^{\ast}}{e}
\left(\frac{n_0}{n_0^{\ast}}-1 \right)^{7/2}. 
\label{nifff}
\end{eqnarray}
The straight lines in Fig. \ref{scale} (b) with slope 7/2 appear to describe
the behaviour of the simulation points quite well. Here we have used
adjustment factors $f_2=2$ for case 1 and $f_2=0.4$ for case 2.

\section{Conclusions}

In summary, we have investigated ion trapping and acceleration 
near the threshold where ion trapping initiates. 
The dynamics of self-regulating 
ion trapping has been identified
as a process of essentially one-dimensional ion wave breaking. 
We have succeeded to determine the power-law profiles of the ion flow
at the instant of wave breaking and the finite amount of charge 
that is trapped and accelerated.
The threshold electric field for ion wave breaking, $E_{\rm{IWB}}$,
has been found to be a function of laser front velocity. 
The trapped ion charge depends on 
the strength of ion wave breaking, which is characterized by 
how much the maximum of the charge-separation field $E_{\rm{max}}$ 
driven by the laser pulse exceeds $E_{\rm{IWB}}$. 
This can be controlled by tuning laser intensity and plasma density. 
Near the threshold this charge is small and localized 
such that high-quality ion bunches with low energy spread 
and beam emittance are expected. 
We hope that the present results stimulate experiments 
to explore this regime, 
which occurs in relativistically transparent plasma  
just below the regime of hole boring.

\section*{Acknowledgements}

B. Liu acknowledges support from the Alexander von Humboldt Foundation.
B. Liu and H. Ruhl acknowledge 
support by the Gauss Centre for Supercomputing (GCS)
Large-Scale Project (Project Nos. pr92na, pr74si), the 
Cluster-of-Excellence Munich Centre for Advanced Photonics (MAP),
and the Arnold Sommerfeld Center (ASC) at Ludwig-Maximilians University of Munich (LMU). 
B. Liu also thanks J. Schreiber for providing 
computing resources on the HYDRA supercomputer at Max-Planck-Institute 
f\"{u}r Quantenoptik (MPQ).

%%%%  figures  %%%%%

%%%%  end of figures  %%%%


\begin{thebibliography}{99}
\bibitem{footnote0}
In Ref. \cite{Akhiezer}, unmagnetized Maxwell equations are used. 
In our case, Eqs. (2,4,5) are physically equivalent to the
unmagnetized Maxwell equations.
\bibitem{footnote1}
By combining Eqs. (\ref{eqbetae},\ref{eqbetai0},\ref{eqez0},\ref{eqne},\ref{eqni0}), 
two integrals are obtained,
$\beta_{\rm{ph}} \left(\gamma_e \beta_e/\mu +  \gamma_i \beta_i\right)
-\left(\gamma_e/\mu +  \gamma_i\right) = \rm{const1.}$
and 
$\tilde{E}^2/2 + 
\beta_{\rm{ph}} \left( \gamma_e \beta_e/\mu +\gamma_i \beta_i \right) 
= \rm{const2.}$
By applying them at the point where 
$\beta_e=\beta_{e,\rm{max}}=\beta_{\rm{ph}}$, $\beta_i=\beta_{i,\rm{min}}$, 
and $\tilde{E}=0$,  
and the point where 
$\beta_e=\beta_i=0$, $\tilde{E}=\tilde{E}_{\rm{max}}=\tilde{E}_{\rm{EWB}}$, 
one has 
$\gamma_{i,m} = \gamma_{\rm{ph}}^2\left(C_1-\beta_{\rm{ph}} \sqrt{C_1^2- 1/\gamma_{\rm{ph}}^2}\right)$,
$\tilde{E}_{\rm{EWB}} = \sqrt{ 2(\gamma_{\rm{ph}}-1)\left(1+ C_2 \right)/\mu },  
$
where 
$\gamma_{i,m}=1/\sqrt{1-\beta_{i,\rm{min}}^2}$, 
$
C_1 = 1+1/\mu-1/(\mu \gamma_{\rm{ph}}), 
$
and 
$
C_2 =  \mu (\gamma_{i,m}-1)/(\gamma_{\rm{ph}}-1). 
$
When $\mu \gg 1$, one gets $C_1=1$, 
$\gamma_{i,m}=1$ and $C_2=0$. Then Eq. (\ref{eewb}) is obtained.
\bibitem{footnote2}
In the hole-boring regime \cite{hb}, all ions in the laser focus are accelerated, 
therefore, the condition of relativistic transparency 
discussed in Ref. \cite{trans} breaks down, 
and the plasma becomes opaque.
\bibitem{footnote3}
In a reflection process 
one has $\mathcal{J}_i'+ \mathcal{J}_{\rm{trap}} = 0 $,  
where $\mathcal{J}_i'$ and $\mathcal{J}_{\rm{trap}}$ 
denote the incoming and reflected current densities  
in the frame comoving with the laser front, respectively. 
It is valid in the region where  $\mathcal{J}_{\rm{trap}} \neq 0$. 
By using the Lorentz transformation one gets 
$\mathcal{J}_i'= \gamma_f(J_i'- v_f q_i n_i')$ 
and  $\mathcal{J}_{\rm{trap}}=\gamma_f (J_{\rm{trap}}- v_f q_i  n_{\rm{trap}})$,
where $J_i'=q_i v_i' n_i'$ and $J_{\rm{trap}}=q_i v_{\rm{trap}} n_{\rm{trap}}$ 
are current densities in the laboratory frame. 
Then the equation 
$\tilde{n}_i' (\beta_i' -\beta_f)
+\tilde{n}_{\rm{trap}}  (\beta_{\rm{trap}} -\beta_f)=0$
is obtained.
\bibitem{footnote4}
The coefficients in Eqs. (\ref{ezrscale}-\ref{ninrscale}) and Eqs. (\ref{ezscale}-\ref{niscale}) may change 
if this assumption breaks down, however, it is easy to find that 
the power-law exponents in these equations are still the same.
\end{thebibliography}
\end{document}